\documentclass[aps,preprint]{revtex4}%
\usepackage{amsfonts}
\usepackage{amsmath}
\usepackage{amssymb}
\usepackage{graphicx}%
\providecommand{\U}[1]{\protect \rule{.1in}{.1in}}

\begin{document}
\title{Polaron formation in the vicinity of a narrow Feshbach resonance in atomic and
exciton-polariton condensates.}
\author{W. Casteels$^{1}$ and M. Wouters$^{1}$}
\affiliation{$^{1}$TQC, Universiteit Antwerpen, Universiteitsplein 1, 2610 Wilrijk, Belgium}

\begin{abstract}
The polaronic system consisting of an impurity in a dilute Bose-Einstein
condensate is considered in the presence of a narrow Feshbach resonance. For
this purpose a coupled-channel model is used, which at the mean field level
predicts the formation of quasiparticles that are a superposition of the
impurity and the molecular states. The impurity-boson interactions and the
coupling between the open and closed channels are then considered weak and a
perturbative calculation of the corrections to the mean field results is
presented. This allows to examine the properties of the quasiparticles, such
as the lifetime and the effective mass. The model is applied to two physical
systems: an impurity atom in a Bose-condensed atomic gas in 3D and a spin down
lower polariton in a Bose-Einstein condensate of spin up lower polaritons in
2D. The model parameters are linked to the physical parameters by identifying
the low energy $T$-matrix and applying a proper renormalization scheme.

\end{abstract}
\maketitle

\section{Introduction}

In recent years ultracold atomic systems have become increasingly popular as
quantum simulators for many-particle models \cite{RevModPhys.80.885}. These
systems are attractive since they are extremely clean as compared to a typical
solid state experiment and the extreme controllability and tunability. One
can, for example, experimentally vary the interatomic interactions by means of
a Feshbach resonance. This is a scattering resonance which occurs when the
scattering energy approaches the binding energy of a molecular state (see for
example Ref. \cite{RevModPhys.82.1225} for a review on Feshbach resonances in
ultracold gases). If the molecule and the scattering atoms have a different
magnetic moment, the difference in energy can be varied with an external
magnetic field. This allows to experimentally tune the interatomic
interactions as a function of an external magnetic field. A distinction is
typically made between a broad or entrance channel dominated resonance and a
narrow or closed channel dominated resonance \cite{RevModPhys.82.1225}. A
broad resonance can be well-described by a single channel model with an
effective interaction amplitude while for the description of a narrow
resonance the molecular state has to be described explicitly with a
coupled-channel model.

Recently the principle of quantum simulation with ultracold gases has been
considered for the Fr\"{o}hlich polaron \cite{PhysRevA.73.043608,
PhysRevLett.96.210401, PhysRevB.80.184504}. The polaron is well-known from
solid state physics for the description of a charge carrier in a charged
lattice, described by the Fr\"{o}hlich Hamiltonian. If the Bogoliubov
approximation is valid the Hamilonian of an impurity in a Bose-Einstein
condensation can be mapped onto the Fr\"{o}hlich Hamiltonian
\cite{PhysRevA.73.043608, PhysRevLett.96.210401, PhysRevB.80.184504}. This
set-up is particularly attractive since the polaronic coupling parameter
depends on the interatomic interaction strength which is tunable with a
Feshbach resonance, whereas in the solid state context this is a material
constant. An example of a promising experiment to probe the polaronic strong
coupling regime is the doping of an ultracold Rb gas with single Cs impurity
atoms from Ref. \cite{PhysRevLett.109.235301} since it exhibits a broad
interspecies Feshbach resonances \cite{PhysRevA.85.032506}. On the other hand
there are also binary systems that exhibit only narrow interspecies Feshbach
resonances, as for example a $^{6}$Li-Na mixture \cite{PhysRevA.85.042721}, in
which case the influence of the resonance on the polaronic properties is not clear.

A related research field that has attracted a great deal of interest in the
context of quantum simulation are quantum fluids of light (See for example
Ref. \cite{RevModPhys.85.299} for a review). If a semiconductor is placed in a
cavity with a strong coupling between the cavity modes and the excitons they
combine to form new quasiparticles, known as polaritons. The coupling strength
is given by the Rabi frequency. By using a cavity with a position dependent
thickness it is possible to experimentally tune the exciton-photon detuning by
probing a different area of the cavity. Since the effective mass of the
polaritons is several orders of magnitude smaller than the exciton mass, they
form a Bose-Einstein condensate at much higher temperatures. The cavity
inevitably exhibits losses, which means that to accomplish a steady state the
system has to be constantly pumped and is never at equilibrium. However, if
the loss rate is sufficiently slow as compared to the thermalization time of
the polariton gas a quasithermal equilibrium can be achieved. By introducing
two different circularly polarized cavity modes which couple to excitons with
different spins it is possible to create spinor polaritons with two spin
states. By using different light intensities for the polarizations a
population imbalance can be introduced between the two spin states which in
the extreme case results in a single spin down lower polariton in a sea of
spin up lower polaritons. If the spin up lower polaritons form a dilute
Bose-Einstein condensate the resulting Hamiltonian can, in the same manner as
for the ultracold gas system, be mapped onto the Fr\"{o}hlich polaron
Hamiltonian. However with the distinction that due to the cavity losses the
system is not at equilibrium. Since two excitons can combine to form a
biexciton also these systems exhibit Feshbach resonances which have been
theoretically predicted in Refs. \cite{PhysRevB.76.045319,
0295-5075-90-3-37001} and confirmed experimentally in\ Ref. \cite{PFR}.

In this paper we are particularly interested in the polaronic system in the
presence of a narrow Feshbach resonance in the two physical contexts outlined
above. Both systems are considered at equilibrium which is expected to be a
good description for an ultracold atomic gas but may be a rather crude
approximation for current experiments on the polaritonic system. A narrow
Feshbach resonance corresponds to weak coupling to the molecular channel so
that perturbation theory is applicable to describe the corrections to the mean
field results. The experimental relevance of this regime is demonstrated by
the wide range of resonance widths that have been observed with ultracold
gases, including narrow resonances \cite{RevModPhys.82.1225} and the recent
observations of a narrow polaritonic Feshbach resonance \cite{PFR}.

The structure of this manuscript is as follows. First the model Hamiltonian
describing two types (denoted as spin up and down) of particles that can form
a molecule is introduced. The manifestation of a Feshbach resnonance within
this model is investigated by calculating the two-body $T$-matrix and a
renormalization scheme is discussed. Then, the spin up particles are assumed
to form a dilute Bose-Einstein condensate by applying the Bogoliubov
approximation. The mean field part of the Hamiltonian is then diagonalized
with another Bogoliubov transformation resulting in two types of
quasiparticles consisting of a superposition of the spin down particles and
the molecules. The interactions are then assumed to be weak and perturbation
theory is used to calculate the lowest order corrections to the self energies
of the quasiparticles. The model is then applied to two physical systems: an
impurity atom in a Bose-condensed atomic gas in 3D and a spin down lower
polariton in a Bose-Einstein condensation of spin up lower polaritons in 2D.
The results are numerically studied for these two systems and finally the
conclusions are presented.

\section{The model system}

The following Hamiltonian is considered for the model system:%
\begin{align}
\widehat{H}  &  =\sum_{\vec{k}}\sum_{\sigma=\uparrow,\downarrow}\xi_{\vec
{k},\sigma}\widehat{a}_{\vec{k},\sigma}^{\dag}\widehat{a}_{\vec{k},\sigma
}+\sum_{\vec{k}}\varepsilon_{\vec{k}}^{M}\widehat{b}_{\vec{k}}^{\dag}%
\widehat{b}_{\vec{k}}\nonumber \\
&  +g\sum_{\vec{k},\vec{q}}\left(  \widehat{b}_{\vec{k}}^{\dag}\widehat
{a}_{\frac{\vec{k}}{2}+\vec{q},\uparrow}\widehat{a}_{\frac{\vec{k}}{2}-\vec
{q},\downarrow}+h.c.\right)  +\frac{U^{\uparrow \uparrow}}{2}\sum_{\vec{k}%
,\vec{k}^{\prime},\vec{q}}\widehat{a}_{\vec{k}+\vec{q},\uparrow}^{\dag
}\widehat{a}_{\vec{k}^{\prime}-\vec{q},\uparrow}^{\dag}\widehat{a}_{\vec
{k}^{\prime},\uparrow}\widehat{a}_{\vec{k},\uparrow}\nonumber \\
&  +U^{\uparrow \downarrow}\sum_{\vec{k},\vec{k}^{\prime},\vec{q}}\widehat
{a}_{\vec{k}+\vec{q},\uparrow}^{\dag}\widehat{a}_{\vec{k}^{\prime}-\vec
{q},\downarrow}^{\dag}\widehat{a}_{\vec{k}^{\prime},\downarrow}\widehat
{a}_{\vec{k},\uparrow}. \label{Ham}%
\end{align}
The first term represents the kinetic energy of the spin up ($\sigma=\uparrow
$) and spin down ($\sigma=\downarrow$) partices with $\widehat{a}_{\vec
{k},\sigma}^{\dag}$ ($\widehat{a}_{\vec{k},\sigma}$) the creation
(annihilation) operators for a particle with wave vector $\vec{k}$, spin
$\sigma$ and $\xi_{\vec{k},\sigma}=E_{\vec{k},\sigma}-\mu_{\sigma}$, where
$\mu_{\sigma}$ is the chemical potential and $E_{\vec{k},\sigma}=E_{0,\sigma
}+\hbar^{2}k^{2}/\left(  2m_{\sigma}\right)  $ the energy with $m_{\sigma}$
the mass. The second term gives the energy of the free molecules, consisting
of a bound state of a spin up and a spin down particle, with $\widehat
{b}_{\vec{k}}^{\dag}$ ($\widehat{b}_{\vec{k}}$) the creation (annihilation)
operators for a molecule with wave vector $\vec{k}$ and $\varepsilon_{\vec{k}%
}^{M}=-\varepsilon^{M}+\hbar^{2}k^{2}/\left(  2M\right)  -\mu_{\uparrow}%
-\mu_{\downarrow}$, with $\varepsilon^{M}$ the binding energy and $M$ the
mass. The first term on the second line represents the formation and
dissociation of the molecule with interaction amplitude $g$. The next term is
the interaction energy of the spin up particles interacting with a contact
potential with amplitude $U^{\uparrow \uparrow}$ and the last term represents
the interaction between the spin up and spin down particles for which also a
contact potential is considered, with amplitude $U^{\uparrow \downarrow}$. Note
that no interactions between the spin-down particles are considered.

\subsection{Two body $T$-matrix and renormalization}

Only considering one spin up and one spin down particle allows a calculation
of the two-body $T$-matrix of the model system with a derivation along the
lines of Ref. \cite{PhysRevA.65.053617}. This results in the following
expression:%
\begin{equation}
\frac{1}{T^{\uparrow \downarrow}\left(  E\right)  }=\left(  U^{\uparrow
\downarrow}+\frac{\left \vert g\right \vert ^{2}}{E-\varepsilon^{M}}\right)
^{-1}-\sum_{\vec{q}}\frac{1}{E-\frac{\hbar^{2}q^{2}}{2m_{r}}+i\delta};
\label{TMAt1}%
\end{equation}
with $E$ the scattering energy, $\delta$ a positive infinitesimal and $m_{r}$
the reduced mass: $m_{r}^{-1}=m_{\downarrow}^{-1}+m_{\uparrow}^{-1}$. Note
that if the dimension is larger than $1$ this expression contains an UV
divergence which can be renormalized with the scheme introduced in Ref.
\cite{PhysRevA.65.053617} for a 3 dimensional set-up and later also applied in
Ref. \cite{PhysRevA.68.053603} for a quasi-two-dimensional atomic gas. This
results in the following renormalization scheme to express the two-body model
parameters $U^{\uparrow \downarrow}$, $g$ and $\varepsilon^{M}$ in terms of the
corresponding physical parameters $U_{P}^{\uparrow \downarrow}$, $g_{P}$ and
$\varepsilon_{P}^{M}$:
\begin{subequations}
\label{Ren}%
\begin{align}
U^{\uparrow \downarrow}  &  =\frac{U_{P}^{\uparrow \downarrow}}{1+RU_{P}%
^{\uparrow \downarrow}};\\
g  &  =\frac{g_{P}}{1+RU_{P}^{\uparrow \downarrow}};\\
\varepsilon^{M}  &  =\varepsilon_{P}^{M}+Rgg_{P};
\end{align}
with:%
\end{subequations}
\begin{equation}
R=\sum_{\vec{q}}\frac{1}{E-\frac{\hbar^{2}q^{2}}{2m_{r}}}.
\end{equation}
Expressing the $T$-matrix (\ref{TMAt1}) as a function of the physical
parameters results in an expression that does not contain an UV divergence:%
\begin{equation}
T^{\uparrow \downarrow}\left(  E\right)  ^{-1}=\left(  U_{P}^{\uparrow
\downarrow}+\frac{\left \vert g_{P}\right \vert ^{2}}{E-\varepsilon_{P}^{M}%
}\right)  ^{-1}+i\pi g\left(  E\right)  ; \label{TMat}%
\end{equation}
where $g\left(  E\right)  $ is the density of states for a free gas with mass
$m_{r}$. Note that the $\left \vert g_{P}\right \vert ^{2}/\left(
E-\varepsilon_{P}^{M}\right)  $ term in (\ref{TMat}) describes the Feshbach resonance.

\subsection{Coupled-channel versus single-channel model}

As mentioned in the introduction a distinction is typically made between broad
or entrance channel dominated and narrow or closed channel dominated
resonances for Feshbach resonances \cite{RevModPhys.82.1225}. As the name
suggests the molecular channel is important for the closed channel dominated
resonances and they should be described with a coupled-channel model, as for
example described by Hamiltonian (\ref{Ham}). For a narrow or entrance channel
dominated resonance the occupation of the closed molecular channel is
typically small and a single-channel model can be used with an effective
interaction amplitude $U_{eff}^{\uparrow \downarrow}$ for the description:%
\begin{equation}
U_{eff}^{\uparrow \downarrow}=U_{P}^{\uparrow \downarrow}-\frac{\left \vert
g_{P}\right \vert ^{2}}{\varepsilon_{P}^{M}}. \label{SingChanUeff}%
\end{equation}

In the following a coupled-channel calculation is developed and analyzed in
perturbation theory in $g_{P}$. Its results will be compared to a perturbative
single-channel calculation.

\section{Bogoliubov approximation}

The temperature is assumed to be sufficiently low such that a macroscopic
number of spin up particles occupy the one-particle ground state and thus form
a Bose Einstein condensation. If the gas is dilute, the Bogoliubov
approximation can be used which transforms the Hamiltonian (\ref{Ham}) into
\cite{Pitaevskii}%
\begin{equation}
\widehat{H}=E^{GP}+\widehat{H}^{MF}+\widehat{H}_{I}.
\end{equation}
The first term is the\ well-known Gross-Pitaevskii energy of the condensate
\cite{Pitaevskii}. The second term is the mean field Hamiltonian, which
consists of all quadratic contributions:%
\begin{equation}
\widehat{H}^{MF}=\sum_{\vec{k}}\left(
\begin{array}
[c]{cc}%
\widehat{a}_{\vec{k}}^{\dag} & \widehat{b}_{\vec{k}}^{\dag}%
\end{array}
\right)  \left(
\begin{array}
[c]{cc}%
\xi_{\vec{k},\downarrow}+U^{\uparrow \downarrow}n_{0} & g\sqrt{n_{0}}\\
g\sqrt{n_{0}} & \varepsilon_{\vec{k}}^{M}%
\end{array}
\right)  \left(
\begin{array}
[c]{c}%
\widehat{a}_{\vec{k}}\\
\widehat{b}_{\vec{k}}%
\end{array}
\right)  +\sum_{\vec{k}\neq0}\varepsilon_{\vec{k}}^{Bog}\widehat{\alpha}%
_{\vec{k}}^{\dag}\widehat{\alpha}_{\vec{k}}. \label{HamMF}%
\end{equation}
The condensate density is denoted as $n_{0}$ and the operators $\left \{
\widehat{\alpha}_{\vec{k}}^{\dag}\right \}  $ and $\left \{  \widehat{\alpha
}_{\vec{k}}\right \}  $ create and annihilate the Bogoliubov excitations with
the corresponding Bogoliubov dispersion:%
\begin{equation}
\varepsilon_{\vec{k}}^{Bog}=\sqrt{\left(  E_{\vec{k}}^{\uparrow}%
+U^{\uparrow \uparrow}n_{0}\right)  ^{2}-\left(  U^{\uparrow \uparrow}%
n_{0}\right)  ^{2}}. \label{BogDisp}%
\end{equation}
Where the notation $E_{\vec{k}}^{\sigma}=E_{\vec{k},\sigma}-E_{0,\sigma}$ was
introduced. The operators $\widehat{\alpha}_{\vec{k}}^{\dag}$ and
$\widehat{\alpha}_{\vec{k}}$ are related to the spin up operators with the
Bogoliubov transformation (for $\vec{k}\neq \vec{0}$):%
\begin{align}
\widehat{\alpha}_{\vec{k}}^{\dag}  &  =u_{k}\widehat{a}_{\vec{k},\uparrow
}^{\dag}+v_{k}\widehat{a}_{-\vec{k},\uparrow};\nonumber \\
\widehat{\alpha}_{\vec{k}}  &  =u_{k}\widehat{a}_{\vec{k},\uparrow}%
+v_{k}\widehat{a}_{-\vec{k},\uparrow}^{\dag}. \label{BogTransf}%
\end{align}
The coefficients $u_{k}$ and $v_{k}$ are given by:%
\begin{align}
v_{k}^{2}  &  =\frac{1}{2}\left(  \frac{E_{\vec{k}}^{\uparrow}+U^{\uparrow
\uparrow}n_{0}}{\varepsilon_{\vec{k}}}-1\right)  ;\\
u_{k}^{2}  &  =\frac{1}{2}\left(  \frac{E_{\vec{k}}^{\uparrow}+U^{\uparrow
\uparrow}n_{0}}{\varepsilon_{\vec{k}}}+1\right)  .
\end{align}
For the chemical potential of the condensate the lowest order approximation
$\mu_{\uparrow}=E_{0,\uparrow}+U^{\uparrow \uparrow}n_{0}$ is used. Note that
this induces a shift of the molecular energy $\varepsilon_{\vec{k}}^{M}$. A
typical length scale of the condensate is the healing length, defined as:
$\xi=\hbar \left(  mn_{0}U^{\uparrow \uparrow}\right)  ^{-1/2}$.

The interaction Hamiltonian $\widehat{H}_{I}$ is given by
\begin{equation}
\widehat{H}_{I}=\sum_{\vec{k},\vec{q}\neq0}\left[  \left(  g\widehat{b}%
_{\vec{q}+\vec{k}}^{\dag}\widehat{a}_{\vec{k}}+U^{\uparrow \downarrow}%
\sqrt{n_{0}}\widehat{a}_{\vec{k}+\vec{q}}^{\dag}\widehat{a}_{\vec{k}}\right)
\left(  u_{q}\widehat{\alpha}_{\vec{q}}-v_{q}\widehat{\alpha}_{-\vec{q}}%
^{\dag}\right)  +h.c.\right]  . \label{HamInt}%
\end{equation}
From now on only a single spin down particle is considered which gives for the
corresponding chemical potential $\mu_{\downarrow}=E_{0,\downarrow}$.

\section{Diagonalization of the mean field part}

The mean field Hamiltonian (\ref{HamMF}) can be diagonalized by applying the
following unitary transformation:%
\begin{equation}
\left(
\begin{array}
[c]{c}%
\widehat{a}_{\vec{k}}^{\dag}\\
\widehat{b}_{\vec{k}}^{\dag}%
\end{array}
\right)  =\left(
\begin{array}
[c]{cc}%
\beta_{\vec{k}} & -\alpha_{\vec{k}}\\
\alpha_{\vec{k}} & \beta_{\vec{k}}%
\end{array}
\right)  \left(
\begin{array}
[c]{c}%
\widehat{\Phi}_{\vec{k}}^{\dag}\\
\widehat{\Psi}_{\vec{k}}^{\dag}%
\end{array}
\right)  . \label{TransfMF}%
\end{equation}
with:%
\begin{subequations}
\begin{align}
\beta_{\vec{k}}^{2}  &  =\frac{1}{2}\left(  1+\frac{E_{\vec{k}}^{\downarrow
}+U^{\uparrow \downarrow}n_{\uparrow}-\varepsilon_{\vec{k}}^{M}}{\sqrt{\left(
E_{\vec{k}}^{\downarrow}+U^{\uparrow \downarrow}n_{0}-\varepsilon_{\vec{k}}%
^{M}\right)  ^{2}+4g^{2}n_{0}}}\right)  ;\\
\alpha_{\vec{k}}^{2}  &  =\frac{1}{2}\left(  1-\frac{E_{\vec{k}}^{\downarrow
}+U^{\uparrow \downarrow}n_{\uparrow}-\varepsilon_{\vec{k}}^{M}}{\sqrt{\left(
E_{\vec{k}}^{\downarrow}+U^{\uparrow \downarrow}n_{0}-\varepsilon_{\vec{k}}%
^{M}\right)  ^{2}+4g^{2}n_{0}}}\right)  .
\end{align}
The mean field Hamiltonian (\ref{HamMF}) is then transformed into:%
\end{subequations}
\begin{equation}
\widehat{H}^{MF}=\sum_{\vec{k}}\left(  \varepsilon_{\vec{k}}^{+}\widehat{\Phi
}_{\vec{k}}^{\dag}\widehat{\Phi}_{\vec{k}}+\varepsilon_{\vec{k}}^{-}%
\widehat{\Psi}_{\vec{k}}^{\dag}\widehat{\Psi}_{\vec{k}}+\varepsilon_{\vec{k}%
}^{Bog}\widehat{\alpha}_{\vec{k}}^{\dag}\widehat{\alpha}_{\vec{k}}\right)  .
\end{equation}
This shows the emergence of two quasiparticles that consist of a superposition
the impurity and the molecular state, with dispersions:%
\begin{equation}
\varepsilon_{\vec{k}}^{\pm}=\frac{1}{2}\left(  E_{\vec{k}}^{\downarrow
}+U^{\uparrow \downarrow}n_{0}+\varepsilon_{\vec{k}}^{M}\pm \sqrt{\left(
E_{\vec{k}}^{\downarrow}+U^{\uparrow \downarrow}n_{0}-\varepsilon_{\vec{k}}%
^{M}\right)  ^{2}+4g^{2}n_{0}}\right)  . \label{EnMF}%
\end{equation}

The mean field effective masses $m^{MF\left(  \pm \right)  }$ can be determined
from the behavior of the dispersion at small $k=\left \vert \vec{k}\right \vert
$:%
\begin{equation}
\frac{1}{m^{MF\left(  \pm \right)  }}=\frac{1}{\hbar^{2}}\left.  \frac
{\partial^{2}\varepsilon_{\vec{k}}^{\pm}}{\partial k^{2}}\right \vert
_{k\rightarrow0}. \label{MFMass}%
\end{equation}
Applying the unitary transformation (\ref{TransfMF}) for the interaction
Hamiltonian (\ref{HamInt}) results in:
\begin{equation}
\widehat{H}_{I}=\sum_{\vec{q}\neq0,\vec{k}}\left(  \gamma_{\vec{k}+\vec{q}%
}\widehat{\Phi}_{\vec{k}+\vec{q}}^{\dag}+\sigma_{\vec{k}+\vec{q}}\widehat
{\Psi}_{\vec{k}+\vec{q}}^{\dag}\right)  \left(  \beta_{k}\widehat{\Phi}%
_{\vec{k}}-\alpha_{k}\widehat{\Psi}_{\vec{k}}\right)  \left(  u_{q}%
\widehat{\alpha}_{\vec{q}}-v_{q}\widehat{\alpha}_{-\vec{q}}^{\dag}\right)
+h.c.; \label{HamInteraction}%
\end{equation}
where the following functions were introduced:%
\begin{align}
\gamma_{\vec{k}}  &  =g\alpha_{\vec{k}}+U^{\uparrow \downarrow}\sqrt{n_{0}%
}\beta_{\vec{k}};\\
\sigma_{\vec{k}}  &  =g\beta_{\vec{k}}-U^{\uparrow \downarrow}\sqrt{n_{0}%
}\alpha_{\vec{k}}.
\end{align}

\section{Perturbative calculation of the self energy}

In general the Green's function of a particle with creation (annihilation)
operators $\left \{  \widehat{\Psi}_{\vec{k}}^{\dag}\right \}  $ ($\left \{
\widehat{\Psi}_{\vec{k}}\right \}  $) is defined as \cite{Mahan}:%
\begin{equation}
G\left(  \vec{k},t-t^{\prime}\right)  =-i\left \langle \mathcal{T}\text{
}\widehat{\Psi}_{\vec{k}}\left(  t\right)  \widehat{\Psi}_{\vec{k}}^{\dag
}\left(  t^{\prime}\right)  \right \rangle . \label{GreenDef}%
\end{equation}
Where $\mathcal{T}$ is the time-ordering operator. The frequency
representation of the Green's function is given by the Fourier transform with
respect to the time:%
\begin{equation}
G\left(  \vec{k},\omega \right)  =\int_{-\infty}^{\infty}dte^{i\omega
t}G\left(  \vec{k},t\right)  \label{GreenFT}%
\end{equation}
and can typically be written as a function of the self energy $\Sigma \left(
\vec{k},\omega \right)  $:%
\begin{equation}
G\left(  \vec{k},\omega \right)  =\frac{\hbar}{\hbar \omega-\varepsilon_{\vec
{k}}-\hbar \Sigma \left(  \vec{k},\omega \right)  }. \label{Green}%
\end{equation}

\begin{figure}
[ptb]
\begin{center}
\includegraphics[
natheight=1.313600in,
natwidth=3.653800in,
height=1.4702in,
width=4.0421in
]%
{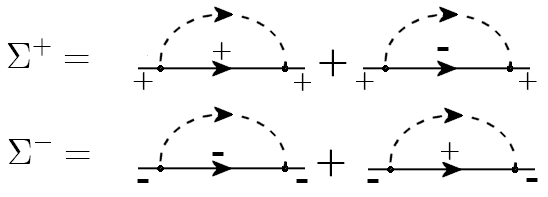}%
\caption{The Feynman diagrams corresponding to the expressions (\ref{SelfEngs}%
) for the self energies of the two branches. The dashed lines denote a
Bogoliubov excitation and the full lines are the unperturbed upper (+) and
lower (-) quasiparticles.}%
\label{Fig: FeynmanGraphs}%
\end{center}
\end{figure}

In appendix \ref{App1} some details are provided on the perturbative
calculation of the lowest non-vanishing contribution of the interaction
Hamiltonian (\ref{HamInt}) to the self energy $\Sigma \left(  \vec{k}%
,\omega \right)  $. The results for the two branches are presented in Fig.
\ref{Fig: FeynmanGraphs} by Feynman diagrams and given by:
\begin{subequations}
\label{SelfEngs}%
\begin{align}
\Sigma^{-}\left(  \vec{k},\omega \right)   &  =\frac{1}{\hbar^{2}}\sum_{\vec
{q}}\left[  \left(  u_{\vec{q}}\sigma_{\vec{k}}\alpha_{\vec{k}-\vec{q}%
}-v_{\vec{q}}\alpha_{\vec{k}}\sigma_{\vec{k}-\vec{q}}\right)  ^{2}G_{0}%
^{-}\left(  \vec{k}-\vec{q},\omega-\omega_{\vec{q}}\right)  \right.
\nonumber \\
&  \left.  +\left(  u_{\vec{q}}\sigma_{\vec{k}}\beta_{\vec{k}-\vec{q}}%
+v_{\vec{q}}\gamma_{\vec{k}-\vec{q}}\alpha_{\vec{k}}\right)  ^{2}G_{0}%
^{+}\left(  \vec{k}-\vec{q},\omega-\omega_{\vec{q}}\right)  \right]
;\label{SelfEnMin}\\
\Sigma^{+}\left(  \vec{K},\omega \right)   &  =\frac{1}{\hbar^{2}}\sum_{\vec
{q}}\left[  \left(  v_{\vec{q}}\beta_{\vec{k}}\gamma_{\vec{k}-\vec{q}}%
-u_{\vec{q}}\gamma_{\vec{k}}\beta_{\vec{k}-\vec{q}}\right)  ^{2}G_{0}%
^{+}\left(  \vec{k}-\vec{q},\omega-\omega_{\vec{q}}\right)  \right.
\nonumber \\
&  \left.  +\left(  u_{\vec{q}}\gamma_{\vec{k}}\alpha_{\vec{k}-\vec{q}%
}+v_{\vec{q}}\sigma_{\vec{k}-\vec{q}}\beta_{\vec{k}}\right)  ^{2}G_{0}%
^{-}\left(  \vec{k}-\vec{q},\omega-\omega_{\vec{q}}\right)  \right]  .
\label{SelfEnPlus}%
\end{align}
These expressions are valid if the following two conditions are satisfied
(with $D$ the dimension):
\end{subequations}
\begin{subequations}
\label{Conditions}%
\begin{align}
\frac{1}{n_{0}\xi^{D}}\left(  \frac{U^{\uparrow \downarrow}}{U^{\uparrow
\uparrow}}\right)  ^{2}  &  \ll1;\label{Cond1}\\
\frac{1}{n_{0}\xi^{D}}\left(  \frac{g}{\sqrt{n_{0}}U^{\uparrow \uparrow}%
}\right)  ^{2}  &  \ll1. \label{Cond2}%
\end{align}
Condition (\ref{Cond1}) shows that the average distance between the particles
should be small as compared to the healing length and the background
interactions where condition (\ref{Cond2}) requires that the coupling to the
molecule should be weak as compared to the condensate interactions. The latter
condition implies that our approximation is only valid for narrow Feshbach
resonances (small $g$).

If there is no coupling with the molecule ($g=0$) the expressions
(\ref{SelfEngs}) reduce to the well-known weak coupling result for the
BEC-impurity Fr\"{o}hlich polaron \cite{0256-307X-26-8-080302,
1751-8121-42-13-135301}:%
\end{subequations}
\begin{equation}
\Sigma^{pol}\left(  \vec{k},\omega \right)  =\frac{n_{0}\left(  U^{\uparrow
\downarrow}\right)  ^{2}}{\hbar}\sum_{\vec{q}}\frac{E_{\vec{q}}^{\uparrow}%
}{\varepsilon_{\vec{q}}^{Bog}}\frac{1}{\hbar \omega-\hbar \omega_{\vec{q}%
}-E_{\vec{k}-\vec{q}}^{\downarrow}+i\delta}. \label{SelfEnPol}%
\end{equation}%
The physical states of the system correspond to the poles of the Green's
function (\ref{Green}). Since the self energy is assumed small with respect to
the mean field dispersion, the argument $\omega$ of the self energy can be
replaced by the unperturbed energy $\varepsilon_{\vec{k}}^{\pm}$ for the
determination of the poles of (\ref{Green}). This results in the following
expression for the energies of the quasiparticles:%

\begin{equation}
E_{\vec{k}}^{\pm}=\varepsilon_{\vec{k}}^{\pm}+\hbar \operatorname{Re}\left[
\Sigma^{\pm}\left(  \vec{k},\varepsilon_{\vec{k}}^{\pm}\right)  \right]  .
\label{PertEn}%
\end{equation}
The inverse of the imaginary part of the self-energy gives the lifetime of the
quasiparticles $\tau_{\vec{k}}^{\pm}$:%
\begin{equation}
\tau_{\vec{k}}^{\pm}=\operatorname{Im}\left[  \Sigma_{0}\left(  \vec
{k},\varepsilon_{\vec{k}}^{\pm}\right)  \right]  ^{-1}. \label{LifeTime}%
\end{equation}
We are also interested in the effective masses $m^{\pm}$ which can be
determined as:%
\begin{align}
\frac{1}{m^{\pm}}  &  =\frac{1}{\hbar^{2}}\left.  \frac{\partial^{2}E_{\vec
{k}}^{\pm}}{\partial k^{2}}\right \vert _{k=0}\label{MassPertCorr}\\
&  =\frac{1}{m^{MF\left(  \pm \right)  }}+\frac{1}{m^{Pert\left(  \pm \right)
}}.
\end{align}
where $m^{MF\left(  \pm \right)  }$ is the mean field effective mass
(\ref{MFMass}) and the second term is the perturbative correction. Since this
correction is assumed to be small we can write:%
\begin{equation}
m^{\pm}\approx m^{MF\left(  \pm \right)  }\left(  1-\frac{m^{MF\left(
\pm \right)  }}{m^{Pert\left(  \pm \right)  }}\right)  .
\end{equation}

\subsection{Renormalization}

A closer look at the arguments of the $\vec{q}$-summations in the expressions
for the self energies (\ref{SelfEngs}) reveals a UV divergency if the
dimension is larger than $1$. This can be cured by using the renormalization
equations (\ref{Ren}). Since only the lowest order contribution of a
perturbation series is considered, the renormalizing equations have to be
considered to the same order. Inserting the renormalization equations
(\ref{Ren}) into the expressions for the mean field energies (\ref{EnMF}) and
only considering the lowest order terms of an expansion in $UR$ and $g^{2}R$
results in:
\begin{subequations}
\label{RenRes}%
\begin{align}
\varepsilon_{\vec{k}}^{-}  &  \rightarrow \varepsilon_{\vec{k}}^{-}%
+\frac{\sigma_{\vec{k}}^{2}}{\hbar^{2}}\sum_{\vec{q}}\frac{1}{\frac{\hbar
^{2}q^{2}}{2m_{r}}-E};\\
\varepsilon_{\vec{k}}^{+}  &  \rightarrow \varepsilon_{\vec{k}}^{+}%
+\frac{\gamma_{\vec{k}}^{2}}{\hbar^{2}}\sum_{\vec{q}}\frac{1}{\frac{\hbar
^{2}q^{2}}{2m_{r}}-E}.
\end{align}
Here, it is understood that the expressions on the right hand side are in
terms of the physical coupling parameters $U_{P}^{\uparrow \downarrow}$,
$g_{P}$ and $\varepsilon_{P}^{M}$. Moreover, since the results for the self
energies (\ref{SelfEngs}) are already of the order we are interested in the
bare coupling parameters can be simply replaced by the physical ones in these
expressions. Using (\ref{RenRes}) for the first term in the expression for the
energies (\ref{PertEn}) exactly cancels the UV divergence.

\section{Physical systems}

The model is now applied to two distinct physical systems. First an impurity
atom in a Bose-condensed atomic gas is considered in a three dimensional
set-up where the impurity can form a molecule with one of the bosons. Then, a
single spin down polariton in the presence of a Bose condensed gas of spin up
polaritons is considered in two dimensions where a bipolariton can be formed
from one spin up and one spin down polariton. In both cases losses are
neglected and the systems are considered at equilibrium.

\subsection{Ultracold atomic Gas system}

The two-body s-wave scattering amplitude $f_{k}$ for the scattering of a spin
up and a spin down atom is related to the $T$-matrix (\ref{TMat}) as
\end{subequations}
\begin{equation}
f_{k}=-\frac{m_{r}}{2\pi \hbar^{2}}T\left(  \frac{\hbar^{2}k^{2}}{2m_{r}%
}\right)  .
\end{equation}
The inverse scattering amplitude is typically expanded for small $k$ as:%
\begin{equation}
-\frac{1}{f_{k}}=\frac{1}{a}+ik-\frac{1}{2}r_{e}k^{2}+\mathcal{O}\left(
k^{4}\right)  ,
\end{equation}
where $a$ is the scattering length, which for the model is given by:%
\begin{equation}
a=\frac{m_{r}}{2\pi \hbar^{2}}\left(  U_{P}^{\uparrow \downarrow}-\frac
{\left \vert g_{P}\right \vert ^{2}}{\varepsilon}\right)  . \label{ScatMod}%
\end{equation}
The effective range $r_{e}$ is related to the Feshbach width $R^{\ast}$ as
$R^{\ast}=-r_{e}/2$, which for the model is:%
\begin{equation}
R^{\ast}=R_{res}^{\ast}\left(  1-\frac{a_{bg}}{a}\right)  ^{2},
\end{equation}
with $R_{res}^{\ast}$ the value at resonance:%
\begin{equation}
R_{res}^{\ast}=\frac{\pi \hbar^{4}}{\left \vert g_{P}\right \vert ^{2}m_{r}^{2}%
V}.
\end{equation}
Note that since our approximation is only valid for small $g_{P}$,
$R_{res}^{\ast}$ should be large. The background statering length $a_{bg}$ is:%
\begin{equation}
a_{bg}=\frac{m_{r}}{2\pi \hbar^{2}}U_{P}^{\uparrow \downarrow}.
\end{equation}
Typically the scattering length in the vicinity of a magnetic Feshbach
resnonace is parametrized as:
\begin{equation}
a=a_{bg}\left(  1-\frac{\Delta B}{B-B_{0}}\right)  , \label{ScatLPhys}%
\end{equation}
with $B$ the magnetic field, $B_{0}$ the location of the resonance and $\Delta
B$ the width, which are experimental parameters. In order to identify the
scattering length of the model (\ref{ScatMod}) with the physical scattering
length (\ref{ScatLPhys}) in the vicinity of the resonance an expansion of the
molecular energy dependence on the magnetic field is performed:%
\[
\varepsilon \left(  B\right)  =\delta \mu \left(  B-B_{0}\right)  +\mathcal{O}%
\left(  \left(  B-B_{0}\right)  ^{2}\right)  ,
\]
where $\delta \mu$ is the difference between the magnetic moment of the
separated atoms and the magnetic moment of the molecular state. This
identification shows that the model parameters can be expressed as a function
of the experimental parameters as follows:%
\begin{align}
U_{P}^{\uparrow \downarrow}  &  =\frac{2\pi \hbar^{2}a_{bg}}{m_{r}};\\
\left \vert g_{P}\right \vert ^{2}  &  =U_{P}^{\uparrow \downarrow}\delta
\mu \Delta B.
\end{align}
Note that $\Delta B$ needs to have the same sign as the background scattering length.

For the Bose gas the amplitude of the contact interactions is given by:%
\begin{equation}
U^{\uparrow \uparrow}=\frac{4\pi \hbar^{2}a_{BB}}{m_{\uparrow}},
\end{equation}
with $a_{BB}$ the boson-boson scattering length. The two conditions
(\ref{Conditions}) for the perturbative result to be valid can be written as a
function of the experimental parameters:%

\begin{subequations}
\label{CondCG}%
\begin{align}
2\pi^{3/2}\sqrt{n_{0}a_{BB}^{3}}\left(  \frac{a_{bg}}{a_{BB}}\right)
^{2}\left(  \frac{m_{\uparrow}}{m_{r}}\right)  ^{2}  &  =\pi \alpha \left(
\frac{m_{\uparrow}}{m_{r}}\right)  ^{2}\ll1\label{CondCG1}\\
4\pi^{3/2}\sqrt{n_{0}a_{BB}^{3}}\frac{m_{\uparrow}}{m_{r}}\frac{a_{bg}}%
{a_{BB}}\frac{\delta \mu \Delta B}{n_{0}U^{\uparrow \uparrow}}  &  \ll1.
\label{CondCG2}%
\end{align}
where the first condition was also expressed as a function of the
dimensionless polaronic coupling parameter $\alpha=a_{bg}^{2}/\left(
a_{BB}\xi \right)  $ \cite{PhysRevB.80.184504}. This shows that the first
condition (\ref{CondCG1}) expresses the diluteness condition $n_{0}a_{BB}%
^{3}\ll1$, weak background interactions and\ that the masses should not be to
different. The second condition expresses the same and additionally that the
width of the resonance $\Delta B$ should be small, corresponding to a narrow
Feshbach resonance.

Since ultralow temperatures are considered the limit $E\rightarrow0$ is taken
in the renormalization expressions (\ref{Ren}).

\subsection{Polaritonic system}

If the exciton-photon coupling is sufficiently strong and the temperature
sufficiently low only the lower polaritons are populated of which the
dispersion is given by \cite{RevModPhys.85.299}:%
\end{subequations}
\begin{equation}
E_{\vec{k}}=\frac{1}{2}\left(  \delta-\sqrt{\delta^{2}+\Omega_{R}^{2}}\right)
+\frac{k^{2}}{2m_{P}}.
\end{equation}
Here $\Omega_{R}$ is the Rabi freqeuncy, $\delta$ is the detuning and $m_{P}$
is the effective mass of the lower polaritons:%
\begin{equation}
m_{P}=\frac{m_{cav}}{\sin^{2}\theta}. \label{PolaritonMass}%
\end{equation}
With $m_{cav}$ the effective photon mass and $\sin \theta$ the photonic
Hopfield coefficient (see later). The interaction parameters depent on the
excitonic Hopfield coefficient $\cos \theta$ as:%
\begin{align}
U^{\uparrow \uparrow}  &  =V\cos^{4}\theta;\\
U_{P}^{\uparrow \downarrow}  &  =V_{bg}\cos^{4}\theta;\\
g_{P}  &  =V_{BX}\cos^{2}\theta.
\end{align}
Where $V$ is the interaction amplitude between the spin up excitons, $V_{bg}$
is the background interaction amplitude between spin up and spin down excitons
and $V_{BX}$ denotes the coupling of the excitons to the biexciton. For the
Hopfield factors the dependence on the wave vector is neglected and they are
given by:%
\begin{align}
\cos \theta &  =\sqrt{\frac{1}{2}\left(  1+\frac{\delta}{\sqrt{\delta^{2}%
+\hbar^{2}\Omega_{R}^{2}}}\right)  };\\
\sin \theta &  =\sqrt{\frac{1}{2}\left(  1-\frac{\delta}{\sqrt{\delta^{2}%
+\hbar^{2}\Omega_{R}^{2}}}\right)  }.
\end{align}
The validity conditions (\ref{Conditions}) can for this system be written as
\begin{subequations}
\label{ConditionsPol}%
\begin{align}
\frac{\cos^{4}\theta}{\sin^{2}\theta}\frac{m_{cav}}{\hbar^{2}}\frac{V_{bg}%
^{2}}{V}  &  \ll1;\\
\frac{1}{\sin^{2}\theta}\frac{m_{cav}}{\hbar^{2}}\frac{V_{BX}^{2}}{n_{0}V}  &
\ll1.
\end{align}
Which again shows that the background interactions and the coupling to the
biexciton should be weak and the Hopfield factor $\sin \theta$ may not be to small.

We are interested in the behavior of the $T$-matrix at low temperature and
thus small scattering energy $E$. In 3D this amounted to simply taking the
limit $E\rightarrow0$ . However in 2D this limit results in a logarithmic IR
divergence. Fortunately, in 2 dimensions the $T$-matrix depends only
logarithmic on the energy \cite{PhysRevA.3.1067}, so that we can set
$E=-\mu_{\uparrow}$ in the renormalizing expressions (\ref{Ren}).

\section{Results and discussion}

In the following we start by summarizing some analytical results for the
background (vanishing coupling to the molecule) and show how the perturbative
single-channel results are obtained. Then the results are examined for the
specific systems of a $^{6}$Li impurity in a Na condensate and for the
polaritonic system in a GaAs-based microcavity.

\subsection{Background results and perturbative single channel-model}

If the coupling to the molecule vanishes ($g=0$), the $\vec{k}$-summation in
the expression for the self energy (\ref{SelfEnPol}) can be performed
analytically in some cases. We start by summarizing these results. We also
indicate how perturbative single-channel results with the effective
interaction strength (\ref{SingChanUeff}) are obtained for the different systems.

\subsubsection{Ultracold atomic Gas system}

If there is no coupling to the molecule ($g=0$) the real part of the impurity
self energy (\ref{SelfEnPol}) at $k=0$ is given by:%
\end{subequations}
\begin{equation}
\frac{\operatorname{Re}\left[  \hbar \Sigma^{bg}\left(  0\right)  \right]
}{\hbar^{2}/\left(  m_{\uparrow}\xi^{2}\right)  }=\frac{1}{2\pi^{2}}\frac
{1}{n_{0}\xi^{3}}\left(  \frac{a_{bg}}{a_{BB}}\right)  ^{2}\frac
{1+\frac{m_{\uparrow}}{m_{\downarrow}}}{1-\frac{m_{\downarrow}}{m_{\uparrow}}%
}\left(  1+\frac{\left(  \frac{m_{\downarrow}}{m_{\uparrow}}\right)  ^{2}%
}{\sqrt{1-\left(  \frac{m_{\downarrow}}{m_{\uparrow}}\right)  ^{2}}}\ln \left[
\frac{m_{\uparrow}-\sqrt{m_{\uparrow}^{2}-m_{\downarrow}^{2}}}{m_{\downarrow}%
}\right]  \right)  . \label{ENPol}%
\end{equation}
The imaginary part of the self energy is only non-zero if $k>k_{c}%
=m_{\downarrow}/\left(  m_{\uparrow}\xi \right)  $. This corresponds to the
Landau criterion for superfluidity \cite{PhysRev.60.356} which states that
energy can only be dissipated to the condensate if $k>k_{c}$. In this case the
imaginary part is given by:%
\begin{equation}
\frac{\operatorname{Im}\Sigma^{bg}\left(  k>k_{c}\right)  }{\hbar/\left(
m_{\uparrow}\xi^{2}\right)  }=-\frac{1}{16\pi}\frac{1}{n_{0}\xi^{3}}\frac
{1}{\xi k}\left(  \frac{a_{bg}}{a_{BB}}\right)  ^{2}\frac{\left(
m_{\downarrow}+m_{\uparrow}\right)  ^{2}}{m_{\downarrow}m_{\uparrow}}\left(
\frac{Q_{c}}{2}\sqrt{4+Q_{c}^{2}}-2\sinh^{-1}\left[  \frac{Q_{c}}{2}\right]
\right)  ; \label{ImPolCG}%
\end{equation}
with:%
\begin{equation}
Q_{c}=\frac{2\frac{m_{\downarrow}}{m_{\uparrow}}\sqrt{1-\left(  \frac
{m_{\downarrow}}{m_{\uparrow}}\right)  ^{2}+\xi^{2}k^{2}}-2\xi k}{\left(
\frac{m_{\downarrow}}{m_{\uparrow}}\right)  ^{2}-1}.
\end{equation}
The effective mass $m_{\downarrow}^{\ast}$ is given by:%
\begin{equation}
\frac{m_{\downarrow}}{m_{\downarrow}^{\ast}}=1+\frac{1}{6\pi^{2}}\frac
{1}{n_{0}\xi^{3}}\left(  \frac{a_{bg}}{a_{BB}}\right)  ^{2}\frac{1}{\left(
1-\frac{m_{\downarrow}}{m_{\uparrow}}\right)  ^{2}}\left(  3+\frac{2+\left(
\frac{m_{\downarrow}}{m_{\uparrow}}\right)  ^{2}}{\sqrt{1-\left(
\frac{m_{\downarrow}}{m_{\uparrow}}\right)  ^{2}}}\ln \left[  \frac
{m_{\uparrow}-\sqrt{m_{\uparrow}^{2}-m_{I}^{2}}}{m_{\downarrow}}\right]
\right)  \label{MassBGCG}%
\end{equation}

The perturbative single-channel model corresponds to these background results
with the replacement of the background scattering length $a_{bg}$ with the
Feshbach enhanced scattering length (\ref{ScatLPhys}).

\subsubsection{Polaritonic system}

If there is no coupling to the molecule ($g=0$) the real part of the self
energy at $k=0$ becomes:%
\begin{equation}
\operatorname{Re}\left[  \hbar \Sigma^{bg}\left(  0\right)  \right]
=\frac{n_{0}V_{bg}^{2}m_{cav}}{4\pi \hbar^{2}}\frac{\cos^{8}\theta}{\sin
^{2}\theta}. \label{EnPolPol}%
\end{equation}
The effective mass in this case is given by:%
\begin{equation}
\frac{1}{m^{\ast}}=\frac{\sin^{2}\theta}{m_{cav}}-\frac{V_{bg}^{2}}{4\pi
\hbar^{2}V}\cos^{4}\theta. \label{EffMassBGPol}%
\end{equation}
For the imaginary part the integral has to be done numerically. Similar as for
the ultracold atomic system, the lifetime is infinite for $k<k_{c}$, with
$k_{c}$ the Landau critical value. For $k>k_{c}$ energy can be dissipated to
the condensate by emitting a Bogoliubov excitation which results in a finite lifetime.

In this case the perturbative single-channel model corresponds to these
background results with the following substution for the background
interaction amplitude $V_{bg}$:%
\begin{equation}
V_{bg}\rightarrow V_{bg}+\frac{\left \vert g\right \vert ^{2}}{2E_{0}%
+\varepsilon^{M}}.
\end{equation}

\subsection{Ultracold atomic Gas system}

The results are now studied for a $^{6}$Li impurity in a Na condensate,
corresponding to the parameters $m_{\downarrow}/m_{\uparrow}=0.263$ and
$a_{bg}/a_{BB}=-1.336$ \cite{PhysRevA.83.042704, PhysRevA.85.042721}. In Ref.
\cite{PhysRevA.85.042721}\ various narrow interspecies Feshbach resonances are
determined for this system. We will not focus on a specific resonance but
examine the typical expected behavior. The difference in magnetic moment
$\delta \mu$ is typically of the order of the Bohr magneton: $\delta \mu
=e\hbar/\left(  2m_{e}\right)  $, with $e$ the elementary charge and $m_{e}$
the electron mass. A coupled-channel calculation in Ref.
\cite{PhysRevA.85.042721}\ revealed that the widths of the resonances $\Delta
B$ are of the order of mG so we take for the width: $\Delta B=-1mG$. For the
Na condensate the typical density $n_{0}=10^{14}$ cm$^{-3}$ is considered.
With these parameters we find:%
\begin{align}
2\pi^{3/2}\left(  n_{0}a_{BB}^{3}\right)  ^{1/2}\left(  \frac{a_{bg}}{a_{BB}%
}\right)  ^{2}\left(  \frac{m_{\uparrow}}{m_{r}}\right)  ^{2}  &
\approx0.25;\\
4\pi^{3/2}\left(  na_{BB}^{3}\right)  ^{1/2}\frac{m_{\uparrow}}{m_{r}}%
\frac{a_{bg}}{a_{BB}}\frac{\delta \mu \Delta B}{n_{0}U^{\uparrow \uparrow}}  &
\approx0.067.
\end{align}
This shows that the conditions (\ref{CondCG}) are satisfied and the system is
in the perturbative regime.

For the units the condensate chemical potential $\mu^{\uparrow}=n_{0}%
U^{\uparrow \uparrow}$, the corresponding healing length $\xi$ and the boson
mass $m_{\uparrow}$ are used. In these units the self energy exhibits a
$\left(  n_{0}\xi^{3}\right)  ^{-1}$ dependence on the density and in the
figures below the self energy is always multiplied with this factor which for
the considered system is $n_{0}\xi^{3}\approx14.5$.

\begin{figure}
[ptb]
\begin{center}
\includegraphics[
height=3.2327in,
width=4.0421in
]%
{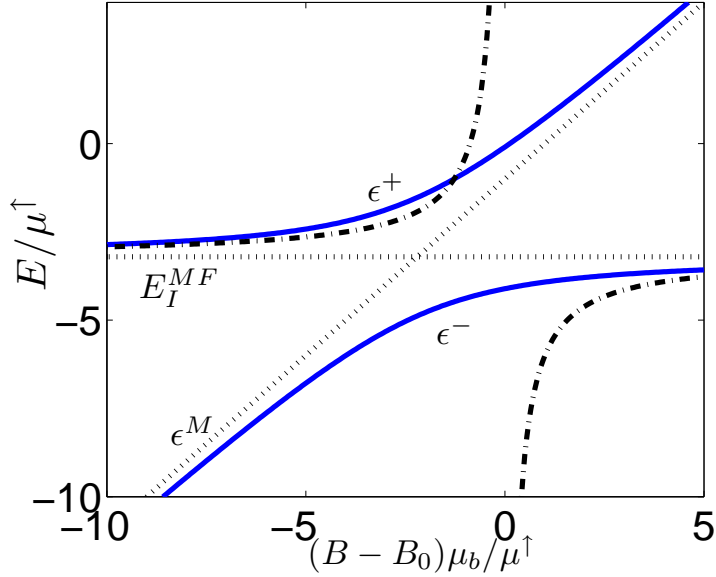}%
\caption{The quasiparticle dispersions $\varepsilon^{\pm}$ at the mean field
level (\ref{EnMF}) due to the mixing of the energy levels of the molecule and
the $^{6}$Li impurity in the presence of a Na condensate as a function of the
magnetic field $B$ at $k=0$. The dotted lines are the energies of the impurity
and the molecule wich are shifted due to the presence of the condensate:
$E_{I}^{MF}=n_{0}U_{P}^{\uparrow \downarrow}$ and $\varepsilon^{M}%
=-\varepsilon_{0}^{M}-n_{0}U^{\uparrow \uparrow}$. The dash-dotted line gives
the single-channel result.}%
\label{Fig: MFSplitUC}%
\end{center}
\end{figure}

In Fig. \ref{Fig: MFSplitUC} the dispersions of the mean field quasiparticles
(\ref{EnMF}) are presented as a function of the applied magnetic field $B$ at
$k=0$. Note that due to the presence of the condensate the resonance is
shifted from $B_{0}$ to%
\begin{equation}
B_{0}^{\ast}=B_{0}+\frac{n_{0}}{\delta \mu}\left(  \frac{2\pi \hbar^{2}a_{bg}%
}{m_{r}}+\frac{4\pi \hbar^{2}a_{BB}}{m_{\uparrow}}\right)  .
\end{equation}
Which for the system under consideration gives $\left(  B_{0}^{\ast}%
-B_{0}\right)  \mu_{b}/\mu^{\uparrow}\approx-2.21$. Also the result of the
perturbative single-channel calculation is presented, which strongly deviates
from the perturbative result around the resonance and diverges at $B=B_{0}$.
The single-channel result provides a good approximation far from the resonance
for the branch with energy closest to the impurity energy (which we will
denote as the impurity-like branch). This could be expected since in this case
the occupation of the closed channel is small.%

\begin{figure}
[ptb]
\begin{center}
\includegraphics[
height=3.2327in,
width=4.0421in
]%
{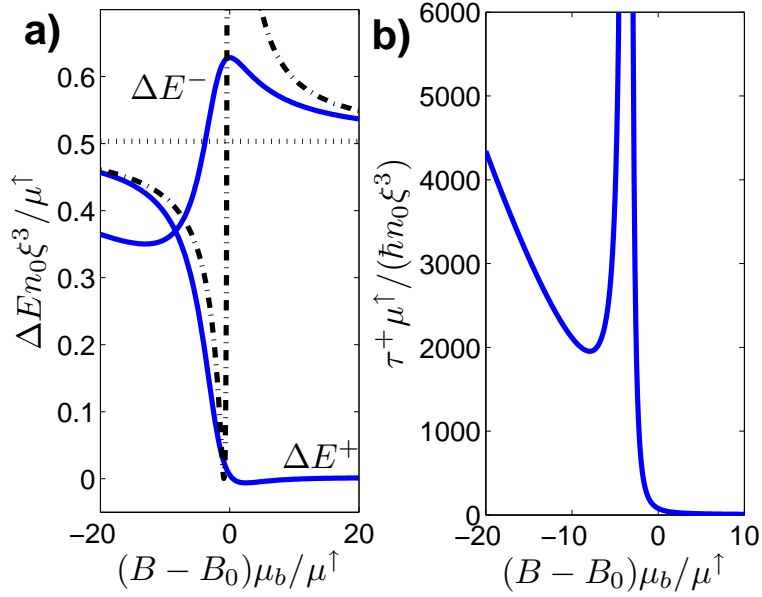}%
\caption{a) The perturbative correction $\Delta E^{\pm}$ to the mean field
quasiparticle energies from Fig. \ref{Fig: MFSplitUC} as a function of the
magnetic field $B$ at $k=0\,$\ for a $^{6}$Li impurity in a Na condensate. The
dashed line is the background result (\ref{ENPol}) and the dot-dashed line is
the perturbative single-channel result. b) The corresponding lifetime of the
upper branch.}%
\label{Fig: EnPertUC}%
\end{center}
\end{figure}

In Fig. \ref{Fig: EnPertUC} (a) the perturbative corrections $\Delta E^{\pm
}=\operatorname{Re}\left[  \hbar \Sigma^{\pm}\left(  0,\varepsilon_{\vec{k}%
}^{\pm}\right)  \right]  $ to the mean field energies, as defined in Eq.
(\ref{PertEn}), are presented as a function of the magnetic field $B$ at
$k=0$. The background result (\ref{ENPol}) and the perturbative single-channel
result are also shown. Again the perturbative single channel result diverges
at $B=B_{0}$ and is only a good approximation for the impurity-like branch far
from the resonance. In Fig. \ref{Fig: EnPertUC} (b) the corresponding lifetime
(\ref{LifeTime}) of the upper branch is presented. The lifetime is infinite in
the limit $B\rightarrow-\infty$ (impurity-like) and zero for $B\rightarrow
\infty$ (molecular-like). In the vicinity of the resonance a strong peak is
observed with a maximum corresponding to an infinite lifetime at $\left(
B-B_{0}\right)  \mu_{b}/\mu^{\uparrow}\approx-3.55$.%

\begin{figure}
[ptb]
\begin{center}
\includegraphics[
height=3.2327in,
width=4.0421in
]%
{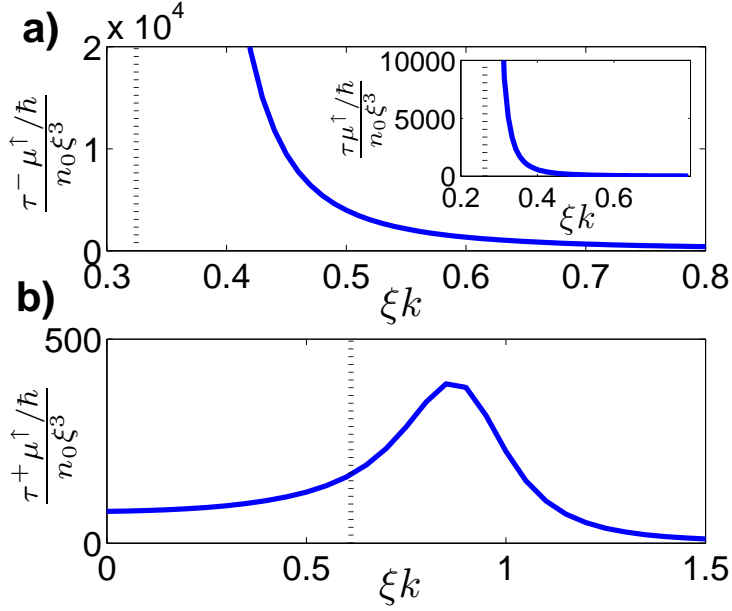}%
\caption{The lifetimes of the lower (a) and the upper (b) quasiparticle
branches are presented as a function of the wave number $k$ for a $^{6}$Li
impurity in a Na condensate. The inset in (a) shows the lifetime\ in the
absence of a molecule. The dashed lines indicate the landau critical values.}%
\label{Fig: TaukCG}%
\end{center}
\end{figure}

The lifetimes of the two branches are presented in Fig. \ref{Fig: TaukCG} as a
function of the wave number $k$ at $B=B_{0}$. Also the lifetime in the absence
of a molecule is presented (the inverse of (\ref{ImPolCG})) which is only
finite for $k>k_{c}$, with $k_{c}$ the Landau critical value, as discussed
before. For the lower branch a similar behavior is observed with an infinite
lifetime at small $k$ untill the Landau critical value $k_{c}^{-}$ is reached.
This value is the lowest possible value of $k$ that satisfies conservation of
energy and momentum for the emission of a Bogoliubov excitation and is thus
determined by:%
\begin{equation}
k_{c}^{\pm}=\min \left \{  k=\left \vert \vec{k}\right \vert :\exists \text{ }%
\vec{q}\text{ }|\varepsilon_{\vec{k}}^{\pm}=\varepsilon_{\vec{k}-\vec{q}}%
^{\pm}+\varepsilon_{\vec{q}}^{Bog}\right \}  .\label{LandauCritVal}%
\end{equation}
For the upper branch a finite lifetime is also found at small $k$ since it can
also decay to the lower branch with the emission of a Bogoliubov excitation.
Once $k$ reaches the Landau critical value $k_{c}^{+}$, as determined by Eq.
(\ref{LandauCritVal}), an extra decay channel opens corresponding to the
emission of a Bogoliubov excitation.

\begin{figure}
[ptb]
\begin{center}
\includegraphics[
height=3.2327in,
width=4.0421in
]%
{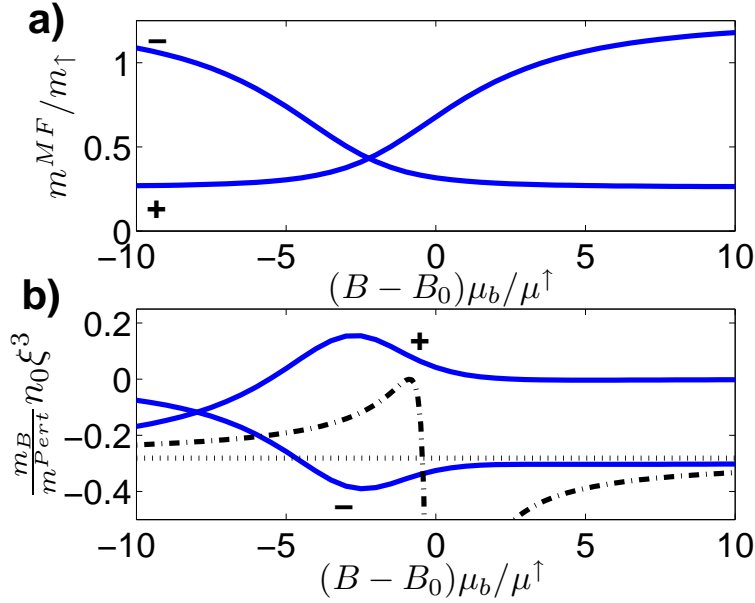}%
\caption{The effective masses are presented for the two branches as a function
of the magnetic field $B$ for a $^{6}$Li impurity in a Na condensate. In (a)
the mean field result (\ref{MFMass}) is presented and in (b) the perturbative
correction to the inverse effective mass is shown. The dotted line is the
background result (\ref{MassBGCG}) and the dash-dotted line is the
perturbative singe-channel result.}%
\label{Fig: MassCG}%
\end{center}
\end{figure}

In Fig. \ref{Fig: MassCG} (a) the mean field effective masses (\ref{MFMass})
are presented for the two quasiparticle branches and in (b) the perturbative
correction $1/m^{Pert\left(  \pm \right)  }$ is shown (as defined in
(\ref{MassPertCorr})). The background result and the perturbative
single-channel result are also depicted. Again we find that the single-channel
result strongly deviates from the perturbative result in the vicinity of the
resonance but is a reasonable approximation for the impurity-like branch far
from the resonance.%

\subsection{Polaritonic system}

The system parameters for a GaAs-based microcavity are introduced
\cite{:/content/aip/journal/apl/65/15/10.1063/1.112877, 2013arXiv1310.6506T,
PFR}, with $\Omega_{R}=3.26$ meV for the Rabi frequency, $E_{bix}=3$ meV for
the molecular binding energy and $m_{cav}=2.62\times10^{-5}m_{e}$ for the
effective photon mass, with $m_{e}$ the electron mass. Only considering the
dominant Coulomb interactions for the spin up lower polaritons and neglecting
the contribution due to phase space philling results in: $V=3a_{B}%
e^{2}/\varepsilon \approx3.73\times10^{-11}$ meV cm$^{2}$ (with $a_{B}$ the
exciton Bohr radius, $e$ the electron charge and $\varepsilon$ the dielectric
constant) \cite{PhysRevB.58.7926}. The ratio of the direct interactions is
approximated with a density independent constant: $V_{bg}/V=-0.39$ and for the
polariton condensate density $n_{0}=10^{10}$ /cm$^{2}$ is considered. For the
coupling strength to the biexciton $V_{BX}=3.73\times10^{-6}$ meV cm is taken.
The biexciton mass is taken infinite since it is much larger than the
effective photon mass. With these parameters we find:
\begin{align}
\frac{\cos^{4}\theta}{\sin^{2}\theta}\frac{m_{cav}}{\hbar^{2}}\frac{V_{bg}%
^{2}}{V}  &  \approx7.56\times10^{-6}\frac{\cos^{4}\theta}{\sin^{2}\theta};\\
\frac{1}{\sin^{2}\theta}\frac{m_{cav}}{\hbar^{2}}\frac{V_{BX}^{2}}{n_{0}V}  &
\approx3.25\times10^{-5}\frac{1}{\sin^{2}\theta}.
\end{align}
This shows that the validity conditions (\ref{ConditionsPol}) are
well-satisfied, except at very small values of $\sin \theta$ (or large positive
values for the detuning $\delta$), which correspond to an almost pure exciton
with large mass (\ref{PolaritonMass}).

For the units the condensate interaction energy at zero detuning $E_{0}%
=Vn_{0}/4$, the corresponding healing length $\xi=\hbar/\sqrt{2m_{cav}E_{0}}$
and twice the photon mass $2m_{cav}$ are used. In these units the self energy
exhibits a $\left(  n_{0}\xi^{2}\right)  ^{-1}$ dependence on the density and
we will multiply the self energy with $n_{0}\xi^{2}$ for the results in this
section. Note that the factor $n_{0}\xi^{2}$ can be quite large and for the
considered system is $n_{0}\xi^{2}\approx1560$.

\begin{figure}
[ptb]
\begin{center}
\includegraphics[
height=3.2327in,
width=4.0421in
]%
{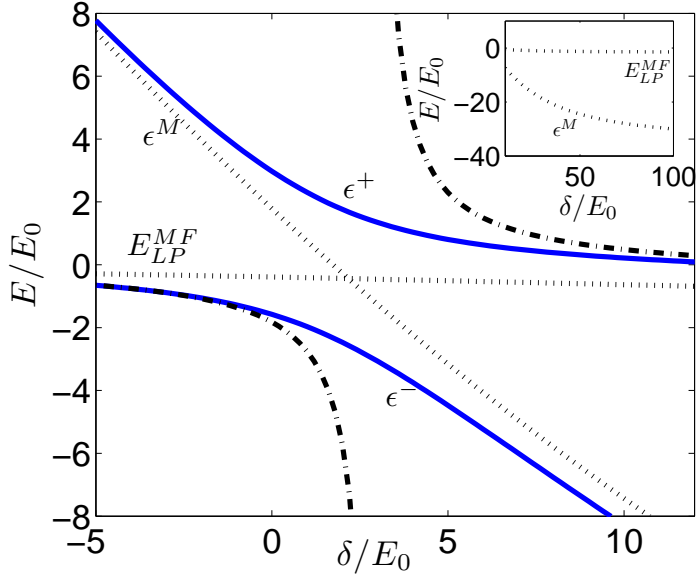}%
\caption{The mean field result for the quasiparticle energies (\ref{EnMF}) as
a function of the detuning $\delta$ for a spin down lower polariton in the
presence of a Bose-condensed gas of spin up lower polaritons in a GaAs-based
microcavity. The dotted lines are the mean field energies with a shift due to
the presence of the condensate of the spin down lower polariton and the
bipolariton: $E_{LP}^{MF}=V_{bg}n_{0}\cos^{4}\theta$ and $\varepsilon
^{M}=-\varepsilon_{0}^{M}-Vn_{0}\cos^{4}\theta-\left(  \delta-\sqrt{\delta
^{2}+\Omega_{R}^{2}}\right)  $ (The inset shows these values at relatively
large $\delta$, showing that both converge to a finite value for
$\delta \rightarrow \infty$). The dash-dotted line gives the result from the
perturbative single-channel calculation.}%
\label{MFEnergiePol}%
\end{center}
\end{figure}

In Fig. \ref{MFEnergiePol} the mean field quasiparticles dispersions
(\ref{EnMF}) are presented as a function of the detuning for $k=0$. The
resonance is shifted from the two-body result due to the presence of the
condensate to $\delta_{R}$, determined by:%
\begin{equation}
V_{bg}n_{0}\cos^{4}\theta=-\varepsilon_{0}^{M}-Vn_{0}\cos^{4}\theta+\left(
\delta_{R}-\sqrt{\delta_{R}^{2}+\hbar^{2}\Omega_{R}^{2}}\right)  ;
\end{equation}
which for the considered system is $\delta_{R}\approx2.17E_{0}$. The
manifestation of these two branches has been observed experimentally for this
system in Ref. \cite{PFR}. Since both the molecular and the lower polariton
energies remain finite in the limit $\delta \rightarrow \infty$, the
quasiparticles remain superpositions of bound and scattering states. Also the
result from a single channel calculation is presented in Fig.
\ref{MFEnergiePol} which results in a reasonable approximation for the
quasiparticle branch that lies close to the lower polariton far from the
resonance.%

\begin{figure}
[ptb]
\begin{center}
\includegraphics[
height=3.2327in,
width=4.0421in
]%
{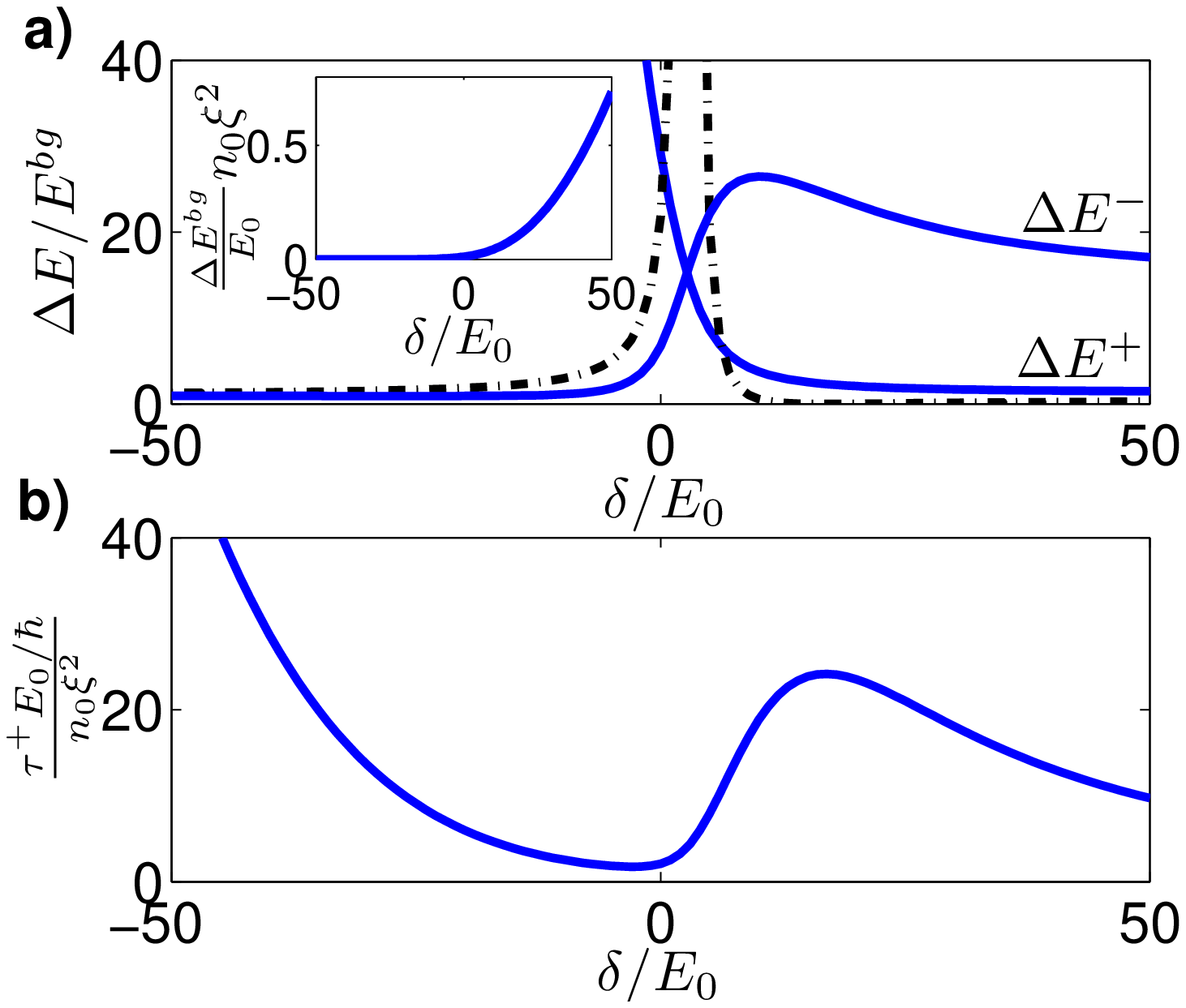}%
\caption{a) The perturbative corrections $\Delta E^{\pm}$ to the mean field
energies divided by the background result (\ref{EnPolPol}) (shown in the
inset) are presented as a function of the detuning $\delta$ for a spin down
lower polariton in the presence of a Bose-condensed gas of spin up lower
polaritons in a GaAs-based microcavity. The dash-dotted line is the result
from a perturbative single-channel calculation. b) The corresponding lifetime
for the upper branch.}%
\label{Fig: EnCorrPol}%
\end{center}
\end{figure}

In Fig. \ref{Fig: EnCorrPol} (a) the perturbative corrections $\Delta E^{\pm
}=\operatorname{Re}\left[  \hbar \Sigma^{\pm}\left(  0,\varepsilon_{0}^{\pm
}\right)  \right]  $, as defined in Eq. (\ref{PertEn}), to the mean field
quasiparticle energies from Fig. \ref{MFEnergiePol} are presented as a
function of the detuning. In order to focus on the effect of the bipolariton
the results are divided by the background result (\ref{EnPolPol}) which is
presented in the inset. Also the perturbative single-channel result is shown
which is a good approximation for the correction to the lower branch at large
negative $\delta$. It is a poorer approximation of the upper branch at large
positive $\delta$ since the quasiparticle remains a superposition in this
limit, as was clear from Fig. \ref{MFEnergiePol}. In Fig. \ref{Fig: EnCorrPol}
(b) the corresponding lifetime is presented for the upper branch. Also here a
peak is observed in the vicinity of the resonance but now with a finite
height. Note furthermore that in contrast with the ultracold atomic system a
finite lifetime is found for $\delta \rightarrow \infty$ ("impurity-like") which
is due to the quasiparticle remaining a superposition of molecular and
impurity states and an infinite lifetime is found for $\delta \rightarrow
-\infty$ ("molecular-like").%

\begin{figure}
[ptb]
\begin{center}
\includegraphics[
height=3.2327in,
width=4.0421in
]%
{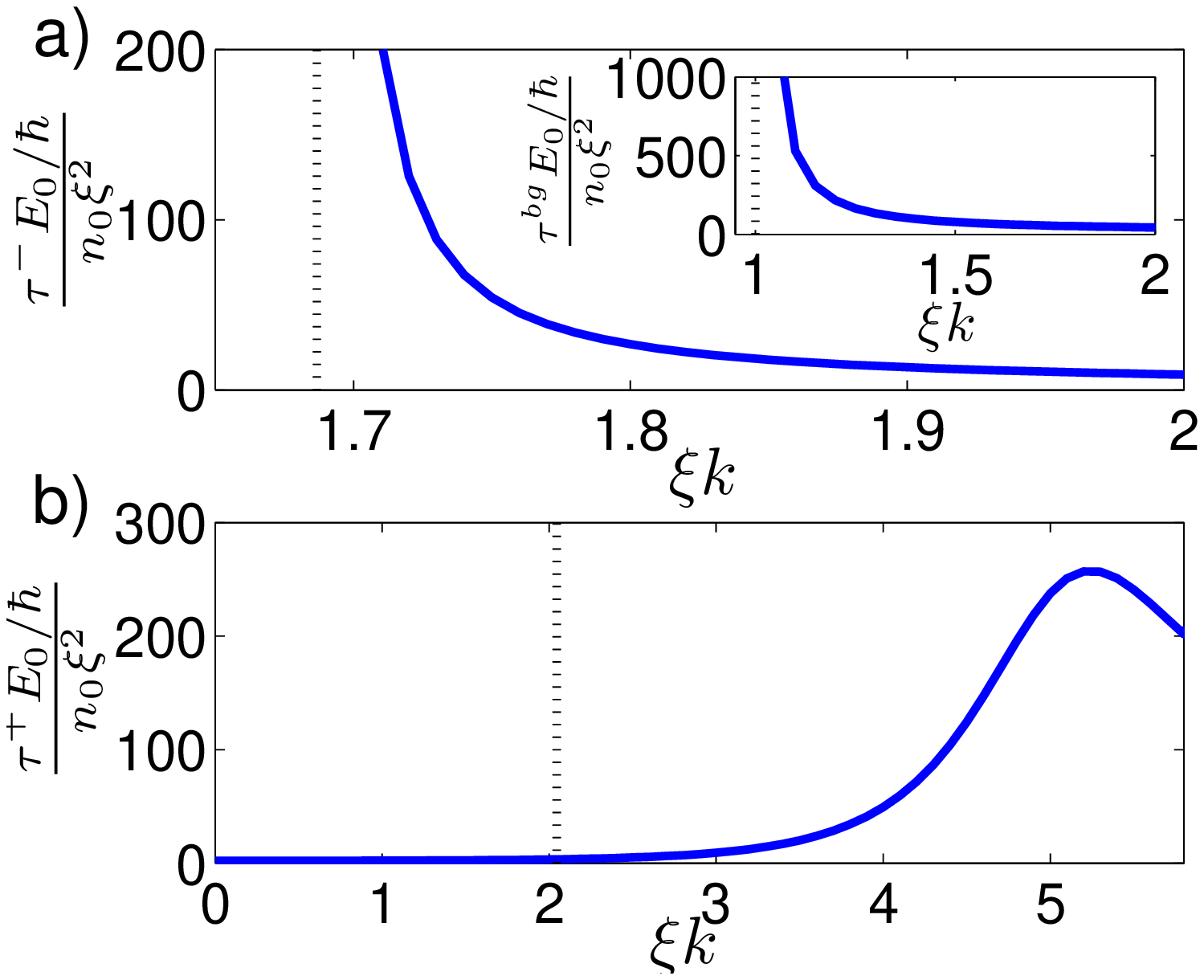}%
\caption{The lifetime of the lower (a) and upper (b) quasiparticle branches
are presented as a function of the wave number $k$ at zero detuning
($\delta=0$) for a spin down lower polariton in the presence of a
Bose-condensed gas of spin up lower polaritons in a GaAs-based microcavity.
The inset in (a) presents the background contribution of the lower polariton
in the absence of a bipolariton. The dotted lines indicate the Landau critical
values, as determined by Eq. (\ref{LandauCritVal}).}%
\label{Fig: LifeTimePol}%
\end{center}
\end{figure}

The lifetimes of the two branches as a function of the wave number $k$ are
presented in Fig. \ref{Fig: LifeTimePol} and the same qualitative behavior is
observed as for the ultracold atomic gas system in Fig. \ref{Fig: TaukCG}. The
background lifetime (shown in the inset of Fig. \ref{Fig: LifeTimePol} (a))
and the lower branch lifetime are infinite for $k<k_{c}$, with $k_{c}$ the
Landau critical value determined by Eq. (\ref{LandauCritVal}), since it is
impossible to dissipate energy to the condensate. The lifetime of the upper
branch is also finite at small $k$ since it can decay to the lower branch with
the emission of a Bogoliubov excitation. Once $k>k_{c}$ the possibility to
emit a Bogoliubov excitation gives an extra contribution to the (inverse)
lifetime.%

\begin{figure}
[ptb]
\begin{center}
\includegraphics[
height=3.2327in,
width=4.0421in
]%
{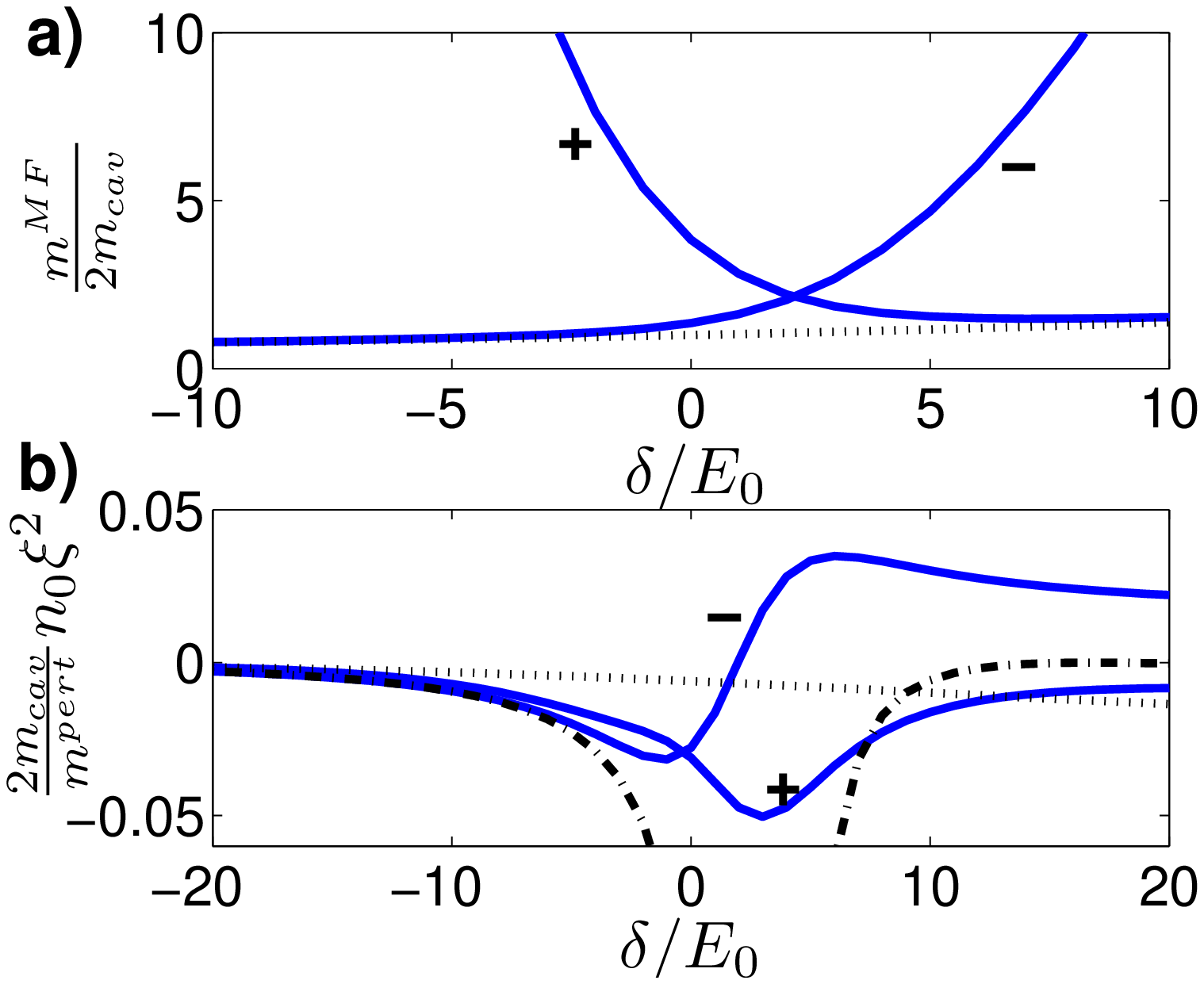}%
\caption{a) The mean field masses of the lower (-) and upper (+) quasiparticle
branches (\ref{MFMass}) as a function of the detuning $\delta$ for a spin down
lower polariton in the presence of a Bose-condensed gas of spin up lower
polaritons in a GaAs-based microcavity. The dotted line presents the lower
polariton mass. b) The corresponding perturbative corrections to the mean
field masses. The dotted line is the background result (\ref{EffMassBGPol})
and the dash-dotted line gives the result from the perturbative single-channel
calculation.}%
\label{Fig: MassPol}%
\end{center}
\end{figure}

The mean field result for the effective masses as a function of the detuning
$\delta$ are presented in Fig. \ref{Fig: MassPol} (a) and the perturbative
correction in (b). The background contribution (\ref{EffMassBGPol}) is also
indicated as well as the result from the perturbative single-channel
calculation. The single-channel result is again a good approximation for the
lower branch for large negative $\delta$ while for large positive $\delta$
there remains a discrepancy since the quasiparticles remain a superposition in
the limit $\delta \rightarrow \infty$.%

\section{Conclusions}

A coupled-channel calculation is presented for the polaronic system consisting
of an impurity in a Bose-condensed gas in the presence of an interspecies
Feshbach resonance. At the mean field level, two quasiparticles branches are
found that consist of a mixing of the impurity and the molecular states. Due
to the presence of the condensate, the location of the Feshbach resonance is
shifted. The corrections to the mean field result are examined within
perturbation theory. This approximation is valid if both the coupling to the
molecular channel and the background interactions are weak with respect to the
condensate interaction energy and if the condensate healing length is
sufficiently large compared to the average interparticle distance.

The model is applied to two physical systems by identifying the low energy
$T$-matrix and using a proper renormalization scheme. This allows to express
the model parameters as a function of the physical parameters. The systems
under consideration\ are a $^{6}$Li impurity atom in a Bose-condensed atomic
Na gas in 3 dimensions and a spin down lower polariton in the presence af a
Bose-Einstein condensate of spin up lower polaritons in a GaAs-based cavity in
two dimensions. These systems are both in the regime where perturbation theory
is valid.

Far from the resonance the results for the quasiparticle branch with an energy
close to the impurity energy are well-reproduced with a perturbative
single-channel calculation. However, in the vicinity of the resonance the
results from the two calculations strongly deviate since the perturbative
single-channel calculation diverges at the resonance while the coupled-channel
calculation predicts finite results.

Some important properties of the quasiparticles, such as the lifetime and the
effective mass were examined. The lifetime of the lower branch is infinite at
small wave number $k$ and becomes finite once $k$ exceeds the Landau critical
value which means it can dissipate energy to the condensate. The lifetime of
the upper branch is also finite at small $k$ since it can decay to the lower
branch with the emission of a Bogoliubov excitation. For $k$ larger than the
Landau critical value it can also emit a Bogoliubov excitation which results
in two contributions to the (inverse) lifetime.

In general the perturbative results are found to be very weak as compared to
the mean field results for the considered systems, in particular for the
polariton system. However, for cold atomic systems, where a wide range in
Feshbach resonance widths exists, the corrections to mean field can be made appreciable.

\appendix{}

\section{Calculation of the self energy\label{App1}}

If the interaction representation is used the Green's function (\ref{GreenDef}%
) can generally be written as a perturbative expansion in the interaction
Hamiltonian $\widehat{H}_{I}$ as \cite{Mahan}:%
\begin{align}
G\left(  \vec{K},t-t^{\prime}\right)   &  =\frac{-i}{\left \langle 0\left \vert
U\left(  \infty,-\infty \right)  \right \vert 0\right \rangle }\sum_{n=0}%
^{\infty}\frac{1}{n!\left(  i\hbar \right)  ^{n}}\nonumber \\
&  \times \int_{-\infty}^{\infty}dt_{1}...dt_{n}\left \langle 0\left \vert
\mathcal{T}\left[  \widehat{\Psi}_{\vec{K}}\left(  t\right)  \widehat{H}%
_{I}\left(  t_{1}\right)  ...\widehat{H}_{I}\left(  t_{n}\right)
\widehat{\Psi}_{\vec{K}}^{\dag}\left(  t^{\prime}\right)  \right]  \right \vert
0\right \rangle , \label{GreenGen}%
\end{align}
where $U\left(  t,t_{0}\right)  $ denotes the time evolution operator:%
\begin{equation}
U\left(  t,t_{0}\right)  =\mathcal{T}\exp \left[  \frac{1}{i\hbar}\int_{t_{0}%
}^{t}dt_{1}\widehat{H}_{I}\left(  t_{1}\right)  \right]  .
\end{equation}
The effect of the denominator $\left \langle 0\left \vert U\left(
\infty,-\infty \right)  \right \vert 0\right \rangle $ is to cancel all
contributions from disconnected diagrams from the series.

For the considered model system with the interaction Hamiltonian
(\ref{HamInteraction}) the first order correction is zero since it involves
the expectation value of a single creation or annihilation operator with
respect to the vacuum. We are interested in the lowest non-vanishing
correction which means we have to calculate the following expectation value:%
\begin{equation}
\left \langle 0\left \vert \mathcal{T}\left[  \widehat{\Psi}_{\vec{k}}\left(
t\right)  \widehat{H}_{I}\left(  t_{1}\right)  \widehat{H}_{I}\left(
t_{2}\right)  \widehat{\Psi}_{\vec{k}}^{\dag}\left(  t^{\prime}\right)
\right]  \right \vert 0\right \rangle .
\end{equation}

\subsection{Expectation values of the quasiparticle operators}

All te needed expectation values are of the following form:%
\begin{equation}
\left \langle 0\left \vert \mathcal{T}\left[
\begin{array}
[c]{c}%
\widehat{\Psi}_{\vec{K}}\left(  t\right)  \left(  \gamma_{\vec{k}_{1}+\vec
{q}_{1}}\widehat{\Phi}_{\vec{k}_{1}+\vec{q}_{1}}^{\dag}\left(  t_{1}\right)
+\sigma_{\vec{k}_{1}+\vec{q}_{1}}\widehat{\Psi}_{\vec{k}_{1}+\vec{q}_{1}%
}^{\dag}\left(  t_{1}\right)  \right)  \left(  \beta_{k_{1}}\widehat{\Phi
}_{\vec{k}_{1}}\left(  t_{1}\right)  -\alpha_{k_{1}}\widehat{\Psi}_{\vec
{k}_{1}}\left(  t_{1}\right)  \right) \\
\times \left(  \gamma_{\vec{k}_{2}+\vec{q}_{2}}\widehat{\Phi}_{\vec{k}_{2}%
+\vec{q}_{2}}^{\dag}\left(  t_{2}\right)  +\sigma_{\vec{k}_{2}+\vec{q}_{2}%
}\widehat{\Psi}_{\vec{k}_{2}+\vec{q}_{2}}^{\dag}\left(  t_{2}\right)  \right)
\left(  \beta_{k_{2}}\widehat{\Phi}_{\vec{k}_{2}}\left(  t_{2}\right)
-\alpha_{k_{2}}\widehat{\Psi}_{\vec{k}_{2}}\left(  t_{2}\right)  \right)
\widehat{\Psi}_{\vec{K}}^{\dag}\left(  t^{\prime}\right)
\end{array}
\right]  \right \vert 0\right \rangle \label{AppExpVal}%
\end{equation}
First the contribution with six creation or annihilation operators for the
lower branch is considered:%
\begin{equation}
\sigma_{\vec{k}_{1}+\vec{q}_{1}}\alpha_{k_{1}}\sigma_{\vec{k}_{2}+\vec{q}_{2}%
}\alpha_{k_{2}}\left \langle 0\left \vert \mathcal{T}\left[  \widehat{\Psi
}_{\vec{K}}\left(  t\right)  \widehat{\Psi}_{\vec{k}_{1}+\vec{q}_{1}}^{\dag
}\left(  t_{1}\right)  \widehat{\Psi}_{\vec{k}_{1}}\left(  t_{1}\right)
\widehat{\Psi}_{\vec{k}_{2}+\vec{q}_{2}}^{\dag}\left(  t_{2}\right)
\widehat{\Psi}_{\vec{k}_{2}}\left(  t_{2}\right)  \widehat{\Psi}_{\vec{K}%
}^{\dag}\left(  t^{\prime}\right)  \right]  \right \vert 0\right \rangle .
\end{equation}
This can be rewritten with Wick's theorem. We will only consider connected
diagrams since the disconnected cancel with the denominator in (\ref{GreenGen}%
) and because we are interested in a single particle at zero temperature, the
expectation value of the occupation number is zero. This means only two
contributions remain:%

\begin{align}
&  \sigma_{\vec{k}_{1}+\vec{q}_{1}}\alpha_{k_{1}}\sigma_{\vec{k}_{2}+\vec
{q}_{2}}\alpha_{k_{2}}\left \langle 0\left \vert \mathcal{T}\left[
\widehat{\Psi}_{\vec{K}}\left(  t\right)  \widehat{\Psi}_{\vec{k}_{1}+\vec
{q}_{1}}^{\dag}\left(  t_{1}\right)  \widehat{\Psi}_{\vec{k}_{1}}\left(
t_{1}\right)  \widehat{\Psi}_{\vec{k}_{2}+\vec{q}_{2}}^{\dag}\left(
t_{2}\right)  \widehat{\Psi}_{\vec{k}_{2}}\left(  t_{2}\right)  \widehat{\Psi
}_{\vec{K}}^{\dag}\left(  t^{\prime}\right)  \right]  \right \vert
0\right \rangle \nonumber \\
&  \rightarrow \sigma_{\vec{K}}\alpha_{\vec{K}}\alpha_{k_{1}}\sigma_{\vec
{k}_{1}}iG_{0}^{-}\left(  \vec{K},t-t_{1}\right)  iG_{0}^{-}\left(  \vec
{k}_{1},t_{1}-t_{2}\right)  iG_{0}^{-}\left(  \vec{K},t_{2}-t^{\prime}\right)
\delta_{\vec{K},\vec{k}_{1}+\vec{q}_{1}}\delta_{\vec{k}_{1},\vec{k}_{2}%
+\vec{q}_{2}}\delta_{\vec{k}_{2},\vec{K}}\nonumber \\
&  +\alpha_{\vec{K}}\sigma_{\vec{K}}\sigma_{\vec{k}_{2}}\alpha_{k_{2}}%
iG_{0}^{-}\left(  \vec{K},t-t_{2}\right)  iG_{0}^{-}\left(  \vec{k}_{2}%
,t_{2}-t_{1}\right)  iG_{0}^{-}\left(  \vec{K},t_{1}-t^{\prime}\right)
\delta_{\vec{K},\vec{k}_{2}+\vec{q}_{2}}\delta_{\vec{k}_{2},\vec{k}_{1}%
+\vec{q}_{1}}\delta_{\vec{k}_{1},\vec{K}},
\end{align}
with $G_{0}^{\pm}\left(  \vec{q},\omega \right)  $ the Green's function of the
unperturbed quasiparticles which in the frequency representation is given by:%
\begin{equation}
G_{0}^{\pm}\left(  \vec{q},\omega \right)  =\frac{\hbar}{\hbar \omega
-\varepsilon_{\vec{k}}^{\pm}+i\delta},
\end{equation}
with $\varepsilon_{\vec{k}}^{\pm}$ the mean field quasiparticle dispersions
(\ref{EnMF}). There are four terms with four creation or annihilation
operators for the lower branch for which Whick's theorem can be used again.
Again we only consider the connected diagrams for a single particle at zero
temperature which results in the following two terms:
\begin{align}
&  \sigma_{\vec{K}}\beta_{k_{1}}\gamma_{\vec{k}_{1}}\alpha_{\vec{K}}iG_{0}%
^{-}\left(  \vec{K},t-t_{1}\right)  iG_{0}^{-}\left(  \vec{K},t_{2}%
-t\prime \right)  iG_{0}^{+}\left(  \vec{k}_{1},t_{1}-t_{2}\right)
\delta_{\vec{K},\vec{k}_{1}+\vec{q}_{1}}\delta_{\vec{k}_{2},\vec{K}}%
\delta_{\vec{k}_{1},\vec{k}_{2}+\vec{q}_{2}}\nonumber \\
&  +\gamma_{\vec{k}_{2}}\alpha_{\vec{K}}\sigma_{\vec{K}}\beta_{k_{2}}%
iG_{0}^{-}\left(  \vec{K},t-t_{2}\right)  iG_{0}^{-}\left(  \vec{K}%
,t_{1}-t\prime \right)  iG_{0}^{+}\left(  \vec{k}_{2},t_{2}-t_{1}\right)
\delta_{\vec{K},\vec{k}_{2}+\vec{q}_{2}}\delta_{\vec{k}_{1},\vec{K}}%
\delta_{\vec{k}_{1}+\vec{q}_{1},\vec{k}_{2}}%
\end{align}
Note that the term with only two creation and annihilation operators of the
lower branch represents a disconnected diagram and thus also cancels with the
denominator of (\ref{GreenGen}), which means we have all the different
contributions for the expectation value (\ref{AppExpVal}).

\subsection{Expectation values of the Bogoliubov operators}

Typically expectation values of the following form are needed:%
\begin{align}
&  \left \langle 0\left \vert \mathcal{T}\left[  \left(  u_{\vec{q}_{1}}%
\widehat{\alpha}_{\vec{q}_{1}}\left(  t_{1}\right)  -v_{\vec{q}_{1}}%
\widehat{\alpha}_{-\vec{q}_{1}}^{\dag}\left(  t_{1}\right)  \right)  \left(
u_{\vec{q}_{2}}\widehat{\alpha}_{\vec{q}_{2}}\left(  t_{2}\right)  -v_{\vec
{q}_{2}}\widehat{\alpha}_{-\vec{q}_{2}}^{\dag}\left(  t_{2}\right)  \right)
\right]  \right \vert 0\right \rangle \nonumber \\
&  =-u_{\vec{q}_{1}}v_{\vec{q}_{1}}\left[  iD\left(  \vec{q}_{1},t_{1}%
-t_{2}\right)  +iD\left(  -\vec{q}_{1},t_{2}-t_{1}\right)  \right]
\delta_{\vec{q}_{1},-\vec{q}_{2}}.
\end{align}
Here $D\left(  \vec{q},t_{1}-t_{2}\right)  $ denotes the unperturbed Green's
function of the Bogoliubov excitations which in frequency representation is
given by:%
\begin{equation}
D\left(  \vec{q},\omega \right)  =\frac{\hbar}{\hbar \omega-\varepsilon_{\vec
{k}}^{Bog}+i\delta},
\end{equation}
with $\varepsilon_{\vec{k}}^{Bog}$ the Bogoliubov dispersion (\ref{BogDisp}).

\subsection{Expression for the self energy}

Bringing everything together gives the following result:%
\begin{align}
&  \int_{-\infty}^{\infty}dt_{1}\int_{-\infty}^{\infty}dt_{2}\left \langle
0\left \vert \mathcal{T}\left[  \widehat{\Psi}_{\vec{K}}\left(  t\right)
\widehat{H}_{I}\left(  t_{1}\right)  \widehat{H}_{I}\left(  t_{2}\right)
\widehat{\Psi}_{\vec{K}}^{\dag}\left(  t^{\prime}\right)  \right]  \right \vert
0\right \rangle \nonumber \\
&  =2\int_{-\infty}^{\infty}dt_{1}\int_{-\infty}^{\infty}dt_{2}\sum_{\vec{q}%
}G_{0}^{-}\left(  \vec{K},t-t_{1}\right)  G_{0}^{-}\left(  \vec{K}%
,t_{2}-t\prime \right) \nonumber \\
&  \times \left \{  \left[  \left(  u_{\vec{q}}\sigma_{\vec{K}}\alpha_{\vec
{K}-\vec{q}}-v_{\vec{q}}\alpha_{\vec{K}}\sigma_{\vec{K}-\vec{q}}\right)
^{2}D\left(  \vec{q},t_{1}-t_{2}\right)  \right.  \right. \nonumber \\
&  \left.  +\left(  v_{\vec{q}}\sigma_{\vec{K}}\alpha_{\vec{K}-\vec{q}%
}-u_{\vec{q}}\alpha_{\vec{K}}\sigma_{\vec{K}-\vec{q}}\right)  ^{2}D\left(
-\vec{q},t_{2}-t_{1}\right)  \right]  G_{0}^{-}\left(  \vec{K}-\vec{q}%
,t_{1}-t_{2}\right) \nonumber \\
&  +\left[  \left(  u_{\vec{q}}\sigma_{\vec{K}}\beta_{\vec{K}-\vec{q}}%
+v_{\vec{q}}\gamma_{\vec{K}-\vec{q}}\alpha_{\vec{K}}\right)  ^{2}D\left(
\vec{q},t_{1}-t_{2}\right)  \right. \nonumber \\
&  \left.  \left.  +\left(  v_{\vec{q}}\sigma_{\vec{K}}\beta_{\vec{K}-\vec{q}%
}+u_{\vec{q}}\gamma_{\vec{K}-\vec{q}}\alpha_{\vec{K}}\right)  ^{2}D\left(
-\vec{q},t_{2}-t_{1}\right)  \right]  G_{0}^{+}\left(  \vec{K}-\vec{q}%
,t_{1}-t_{2}\right)  \right \}
\end{align}

By going to the frequency representation with a Fourier transformation with
respect to the time, as in Eq. (\ref{GreenFT}), the following expression is
retrieved for the second order correction to the Green's function of the lower
quasiparticle branch:%
\begin{align}
G^{-\left(  2\right)  }\left(  \vec{k},\omega \right)   &  =\frac{i}{\hbar^{2}%
}\int_{-\infty}^{\infty}\frac{d\nu}{2\pi}\sum_{\vec{q}}G_{0}^{-}\left(
\vec{K},\omega \right)  G_{0}^{-}\left(  \vec{K},\omega \right) \nonumber \\
&  \times \left \{  \left[  \left(  u_{\vec{q}}\sigma_{\vec{K}}\alpha_{\vec
{K}-\vec{q}}-v_{\vec{q}}\alpha_{\vec{K}}\sigma_{\vec{K}-\vec{q}}\right)
^{2}D\left(  \vec{q},\nu \right)  \right.  \right. \nonumber \\
&  \left.  +\left(  v_{\vec{q}}\sigma_{\vec{K}}\alpha_{\vec{K}-\vec{q}%
}-u_{\vec{q}}\alpha_{\vec{K}}\sigma_{\vec{K}-\vec{q}}\right)  ^{2}D\left(
-\vec{q},-\nu \right)  \right]  G_{0}^{-}\left(  \vec{K}-\vec{q},\omega
-\nu \right) \nonumber \\
&  +\left[  \left(  u_{\vec{q}}\sigma_{\vec{K}}\beta_{\vec{K}-\vec{q}}%
+v_{\vec{q}}\gamma_{\vec{K}-\vec{q}}\alpha_{\vec{K}}\right)  ^{2}D\left(
\vec{q},\nu \right)  \right. \nonumber \\
&  \left.  \left.  +\left(  v_{\vec{q}}\sigma_{\vec{K}}\beta_{\vec{K}-\vec{q}%
}+u_{\vec{q}}\gamma_{\vec{K}-\vec{q}}\alpha_{\vec{K}}\right)  ^{2}D\left(
-\vec{q},-\nu \right)  \right]  G_{0}^{+}\left(  \vec{K}-\vec{q},\omega
-\nu \right)  \right \}
\end{align}
This can be written as a function of the self energy $\Sigma^{-\left(
2\right)  }\left(  \vec{k},\omega \right)  $:%
\begin{equation}
G^{-\left(  2\right)  }\left(  \vec{k},\omega \right)  =G_{0}^{-}\left(
\vec{K},\omega \right)  \Sigma^{-\left(  2\right)  }\left(  \vec{k}%
,\omega \right)  G_{0}^{-}\left(  \vec{K},\omega \right)  ,
\end{equation}
showing that the self energy is given by:%
\begin{align}
\Sigma^{-\left(  2\right)  }\left(  \vec{k},\omega \right)   &  =\frac{i}%
{\hbar^{2}}\int_{-\infty}^{\infty}\frac{d\nu}{2\pi}\sum_{\vec{q}}\left \{
\left[  \left(  u_{\vec{q}}\sigma_{\vec{K}}\alpha_{\vec{K}-\vec{q}}-v_{\vec
{q}}\alpha_{\vec{K}}\sigma_{\vec{K}-\vec{q}}\right)  ^{2}D\left(  \vec{q}%
,\nu \right)  \right.  \right. \nonumber \\
&  \left.  +\left(  v_{\vec{q}}\sigma_{\vec{K}}\alpha_{\vec{K}-\vec{q}%
}-u_{\vec{q}}\alpha_{\vec{K}}\sigma_{\vec{K}-\vec{q}}\right)  ^{2}D\left(
-\vec{q},-\nu \right)  \right]  G_{0}^{-}\left(  \vec{K}-\vec{q},\omega
-\nu \right) \nonumber \\
&  +\left[  \left(  u_{\vec{q}}\sigma_{\vec{K}}\beta_{\vec{K}-\vec{q}}%
+v_{\vec{q}}\gamma_{\vec{K}-\vec{q}}\alpha_{\vec{K}}\right)  ^{2}D\left(
\vec{q},\nu \right)  \right. \nonumber \\
&  \left.  \left.  +\left(  v_{\vec{q}}\sigma_{\vec{K}}\beta_{\vec{K}-\vec{q}%
}+u_{\vec{q}}\gamma_{\vec{K}-\vec{q}}\alpha_{\vec{K}}\right)  ^{2}D\left(
-\vec{q},-\nu \right)  \right]  G_{0}^{+}\left(  \vec{K}-\vec{q},\omega
-\nu \right)  \right \}  . \label{SelfEnApp}%
\end{align}
Using the Dyson equation it is clear that the self energy $\Sigma \left(
\vec{k},\omega \right)  $ from (\ref{Green}) can be approximated by the self
energy (\ref{SelfEnApp}). The frequency integral in Eq. (\ref{SelfEnApp}) can
be done with contour integration which finally results in the expression
(\ref{SelfEnMin}) for the self energy of the lower branch. For the upper
branch a similar calculation can be done which results in expression
(\ref{SelfEnPlus}).

\bibliographystyle{Science}
\bibliography{NarrowFB}

\end{document}